\documentstyle[preprint,tighten,prd,aps,eqsecnum]{revtex}
\begin{document}
\draft
\title{Second post-Newtonian gravitational radiation reaction
for two-body systems: Nonspinning bodies}
\author{A. Gopakumar$^1$, Bala R. Iyer$^1$, and Sai Iyer$^2$} 
\address{$^1$ Raman Research Institute, Bangalore 560 080, India}
\address{$^2$ Physical Research Laboratory, Ahmedabad 380 009, India}
\date{\today}
\maketitle
\begin{abstract}
Starting from the recently obtained 
2PN accurate forms of the energy and angular
momentum fluxes from  inspiralling 
compact binaries, we deduce the gravitational radiation
reaction to 2PN order beyond the quadrupole approximation
-- 4.5PN terms in the equation of motion -- using the refined balance
method proposed by Iyer and Will. We  explore critically
the features of their construction and illustrate them by
contrast to other possible variants. 
The equations of motion 
are valid for general binary orbits and for a class of 
coordinate gauges. The limiting cases of circular orbits
and radial infall are also discussed.
\end{abstract}

\section{Introduction}
\label{sec:intro}

Inspiralling compact binaries are the most promising sources of
gravitational radiation in the near future for ground-based laser
interferometric detectors like LIGO\cite{ligo}, VIRGO\cite{virgo},
GEO600\cite{geo} and TAMA\cite{tama}.
The method of matched filtering will be employed to search for
the inspiral waveforms and extract the information they carry
\cite{kip30,bfs91}. For this method to be successful, one needs to use
templates that are extremely accurate in their description of
the evolution of the orbital phase, which in turn requires a detailed
understanding of how radiation damping (reaction) influences orbital evolution
\cite{curtprl,finncher93,curtena,bsd}.

The idea of a damping force associated with an interaction that propagates
with a finite velocity was first discussed in the context of 
electromagnetism by Lorentz\cite{lor1}. He obtained it by a direct 
calculation of the total force
acting on a small extended particle due to its self-field. The answer was
incorrect by a numerical factor and the correct result was first 
obtained by Planck\cite{planck}  using a `heuristic' argument based on 
energy balance 
which prompted Lorentz\cite{lor2} to re-examine
his self-field calculations and confirm Planck's result, 
\[
F^i\,=\,\frac{2}{3}\,\,\frac{e^2}{c^3}\,\ddot{v}^i\;\;,\;\]
 where $v_i$  is the velocity of the particle.
The  relativistic generalization of the
radiation reaction by Abraham\cite{abraham} based on arguments of energy and 
linear momentum balance  preceded by a few years the
direct relativistic self-field calculation by Schott\cite{schott}
and illustrates
the utility of this heuristic, albeit less rigorous, approach\cite{td82}.

The  argument based on energy balance proceeds thus: 
A non-accelerated particle does not radiate and satisfies
Newton's (conservative) equation of motion. If it
is accelerated, it radiates, loses energy and this implies
damping terms in the equation of motion.
Equating the work done by the reactive force on the particle in
a unit time interval to the negative of the energy radiated by
the accelerated particle in that interval (Larmor's formula)
the reactive acceleration is determined  and one is led to the
Abraham-Lorentz equation of motion for the charged particle.
The direct method  of obtaining  radiation damping, on the other hand, 
is based on the
evaluation of the self-force. Starting with the momentum
conservation law for the electromagnetic fields
one  rewrites this as Newton's equation of motion
by decomposing the electromagnetic fields into an `external field' and a 
`self-field'. Expanding the
self-field in terms of potentials, solving for them in terms of retarded
fields and finally making a retardation expansion, one obtains
the required equation of motion when one goes to the point particle
limit\cite{jackson}.

As in the electromagnetic case, the approach to gravitational radiation
damping has been based on the balance methods, the reaction
potential or a full iteration of Einstein's equation. 
The first computation in general relativity was by Einstein\cite{einstein}  
who derived the loss
in energy of a spinning rod by a far-zone energy flux computation. The
same was derived  by Eddington\cite{eddington} by a direct near-zone radiation
damping approach. He also pointed out that the physical mechanism causing
damping was the effect discussed by Laplace\cite{laplace}, that if 
gravity was not
propagated instantaneously, reactive forces could result. 
An useful development was the introduction of the
radiation reaction potential  by Burke\cite{wb69} and Thorne\cite{kt69} 
using the method
of matched asymptotic expansions. In this approach, 
one  derives the equation of motion by constructing an outgoing wave
solution of Einstein's equation in some convenient gauge and then matching
it to the near-zone solution. Restricting attention  only to   lowest
order Newtonian terms and terms 
sensitive to the outgoing (ingoing) boundary conditions and neglecting all
other terms, one obtains the required result.
The  first complete
direct calculation \`{a} la Lorentz of the gravitational radiation reaction 
force was by Chandrasekhar and Esposito\cite{ce70}.  
Chandrasekhar and collaborators\cite{c69,cn69} 
developed a systematic post-Newtonian expansion for extended perfect 
fluid systems and put together
correctly the necessary elements like the Landau-Lifshiftz pseudotensor, 
the retarded potentials and the near-zone expansion. These works established
the  balance  equations to Newtonian order, albeit for weakly self-gravitating
fluid systems. The revival of interest in these issues following the
discovery of the binary pulsar and the applicability of these very
equations to binary systems of compact objects follows from the works of
Damour\cite{td82,td82a} and
Damour and Deruelle\cite{dd81}.

In the context of the binary pulsar timing, the accuracy reached by
the Newtonian balance equations is  amply adequate. The case
of inspiralling binaries as sources for the interferometric 
gravitational wave detectors is very different. The extremely
high phasing accuracy requirement makes mandatory the control
of reactive terms way beyond the Newtonian. This has prompted
on the one hand, work on generation aspects to compute the
far-zone flux of energy and angular momentum carried by gravitational
waves and on the other, work on the radiation reaction aspects
to compute the effect on the orbital motion of the emission of
gravitational radiation. As in the electromagnetic case, the computation
of the reactive acceleration assuming balance equations is simpler
than the computation of the damping terms by a direct near-field iteration.
The computation of the energy and angular momentum fluxes
at the lowest Newtonian order (quadrupole equation) requires 
the equation of motion at only Newtonian order. {\it Assuming} the
balance equations one can infer the lowest order (2.5PN)
radiation damping whose direct computation, as mentioned
before, requires a 2.5PN iteration of the near-zone equations.  
Similarly, the computation of the 1PN corrections to the lowest order
quadrupole luminosity requires the 1PN accurate
equations of motion, but is potentially equivalent to the
3.5PN terms in the equation of motion.
This motivated Iyer and Will( IW )\cite{iwl,iw95} to propose a refinement 
of the text-book
\cite{landau} treatment of the energy balance method used to discuss 
radiation damping. This generalization
uses both energy and angular momentum balance to deduce the 
radiation reaction force for a binary system made of
nonspinning structureless particles moving on general orbits.
Starting from the 1PN
conserved dynamics of the two-body system, and the radiated
energy and angular momentum in the gravitational waves, 
and taking into account the arbitrariness of the `balance' upto 
total time derivatives, they determined the
2.5PN and 3.5PN terms in the equations of motion of
the binary system. The part not fixed by the balance equations was 
identified with the freedom still residing in the choice of the
coordinate system at that order.
Thus, starting from the far-zone flux formulas, one deduces
a formula that is suitable for evolving general orbits of compact
binaries of arbitrary mass ratio and that includes 1PN corrections to
the dominant Newtonian radiation reaction terms.
Blanchet\cite{lb93,lb96}, on the other hand, obtained the post-Newtonian 
corrections to the radiation reaction force from
first principles using a combination of post-Minkowskian, 
multipolar and post-Newtonian schemes together with techniques of
analytic continuation and asymptotic matching. By looking at
``antisymmetric'' waves -- a solution of the d'Alembertian equation
composed of retarded wave minus advanced wave, regular all over the
source -- and matching, one obtains a radiation reaction tensor potential
that generalizes the Burke-Thorne reaction potential\cite{mtw}, in terms of
explicit integrals over matter fields in the  source. The {\it validity} 
of the balance equations upto 1.5PN is also proved. By specializing
this potential to two-body sytems, 
Iyer and Will\cite{iw95}  checked that this solution indeed corresponds to a 
unique and consistent choice of coordinate system. This provides
a delicate and non-trivial check on the validity of the 1PN reaction
potentials and the overall consistency of the direct methods based
on iteration of the near-field equations and indirect methods based
on energy and angular momentum balance.

As emphasized earlier, much better
approximations are needed to reach the precision of future
gravitational-wave astronomy \cite{curtprl}. In the limit
where one mass is much smaller than the other,
numerical and analytical computations based on black hole
perturbation theory have been performed to 
the 5.5PN order \cite{EP,TN,TS,HT,TSTS,TTHTMS}, a recent
result being the analytical expression to 5.5PN order for the energy
flux from a test particle moving in a circular orbit around a 
Schwarzschild black hole \cite{TTHTMS}.
Ryan\cite{ryan1,ryan2} has investigated the effect of
gravitational radiation reaction, first on circular, and later even for
non-equatorial orbits
around a spinning black hole.
Recently Mino, Sasaki and Tanaka\cite{mst96} have derived the leading order
correction to the equation of motion of a particle which presumably describes
the effect of gravitational radiation reaction by two methods: 
an extension of the Dewitt-Brehme formalism and the method of
asymptotic matching.

On the other hand, for bodies of comparable masses,
recently two independent teams\cite{BDIWW,BIWW,bdi2pn,ww2pn,gi96}
 have derived the 2PN accurate
gravitational waveform and the  associated energy and angular momentum 
fluxes for inspiralling compact binaries through 2PN order by two
independent methods: the BDI approach based on a mixed  multipolar 
post-Minkowskian and post-Newtonian framework together with
asymptotic matching and analytic continuation\cite{bdigen}
 and the recently
improved Epstein-Wagoner (EW) \cite{ew75} formalism by Will and Wiseman 
\cite{ww2pn} which provides a method to carefully handle the divergences
of the older EW treatment. 
In view of the above discussion it is natural to investigate
the possibility of extending the treatment of Iyer and Will
to 2PN accuracy beyond the Newtonian (2.5PN) radiation reaction
and this is what we propose to take up in this paper.
The knowledge of the reactive acceleration beyond the lowest
order could also have practical uses. For instance,
Lincoln and Will\cite{LW90}  have studied the
late-time orbital evolution of compact binaries with arbitrary
mass ratios. They described the orbit using the osculating orbital
elements of celestial mechanics and used the Damour-Deruelle two-body
equations of motion including Newtonian radiation reaction terms
\cite{dd81,td82} to evolve these orbital elements. The extension
of this work to include 1PN radiation reaction is still not available.
 Recently, a 2PN accurate
description for the motion of spinning compact binaries of arbitrary
mass ratio was obtained in a generalized quasi--Keplerian
parameterization initially suggested by Damour and Sch\"{a}fer 
\cite{DS88,SW93,W95,WR95}. These orbital elements have also not been evolved to
2PN radiation reaction order. Our present computation is a step in
that direction. 
These attempts to study the evolution of binary orbits 
would be complementary to those using the test particle limit
\cite{ryan1,ryan2}.

To summarise: Starting from 2PN accurate energy and angular momentum 
fluxes for  compact binaries of arbitrary mass ratio moving in quasi--elliptical
orbits \cite{gi96,ww2pn}, 
 we  obtain the 4.5PN 
reactive terms in the equations of motion by an extension of the IW method. 
Schematically,  the 
equations of motion for spinless bodies of arbitrary mass ratio are
\begin{equation}
\label{a-schem}
{\bf a} \equiv \frac{d^2{\bf x}}{dt^2} 
    \approx -\frac{m{\bf x}}{r^3}
    [1+O(\epsilon)+O(\epsilon^2)+O(\epsilon^{2.5})
     +O(\epsilon^3) +O(\epsilon^{3.5})+
     O(\epsilon^4)+O(\epsilon^{4.5})+\dots]\,,
\end{equation}
where $\bf x$ and $r=\vert\bf x\vert$ denote the separation vector
and distance between the bodies, and $m=m_1 + m_2$ denotes the
total mass.
The quantity $\epsilon$ is a small expansion parameter that
satisfies $\epsilon \sim (v/c)^2 \sim Gm/(rc^2)$,
where $v$ and $r$ are the orbital velocity and separation
of the binary system.
The symbols $O(\epsilon)$ 
and $O(\epsilon^2)$ represent post-Newtonian
(PN),  post-post-Newtonian (2PN) corrections and so on.
Gravitational radiation reaction first appears
at $O(\epsilon^{2.5})$ beyond Newtonian gravitation, or at 2.5PN
order.  We call this the ``Newtonian'' radiation reaction. 
``Post-Newtonian'' radiation reaction terms, at $O(\epsilon^{3.5})$,
were obtained by  Iyer and Will\cite{iwl,iw95} and Blanchet \cite{lb93,lb96}.
Here we obtain the 2PN radiation reaction, at $O(\epsilon^{4.5})$. 
The 4.5PN reactive terms are determined in terms of twelve arbitrary
parameters, which along the lines of
\cite{iwl,iw95}, are associated with the possible 
residual `gauge' choice at the  4.5PN order.
These results valid for general orbits are specialized to the two 
 complementary cases of circular orbits and radial infall.
The expressions for  $\dot r$ and $\dot\omega$  for the quasi--circular
orbits  and $\dot{z}$
for radial infall to 4.5PN order are in agreement with
\cite{BDIWW,SPW95} as required.
We next examine critically the origin of the `redundant' equations
in the formalism and examine our understanding of this
redundancy by exploring variant schemes  which differ from the
original IW scheme in their choice of the functional forms for the arbitrary
terms in energy and angular momentum.  

The paper is organized as follows. In section~\ref{sec:IW}, we
describe the IW method to obtain the 2PN reactive terms.
Section~\ref{sec:altscheme} examines the question of redundant
equations and explores `variants' of the original IW scheme that
differ in their choice of the ambiguities in energy and angular
momentum.  Section~\ref{sec:gauge} discusses the question of the
undetermined parameters and arbitrariness in the choice of the gauge,
in particular at 4.5PN order.  Section~\ref{sec:qcho} is devoted to
the particular cases of quasi-circular orbits and head-on infall.
Section~\ref{sec:conclusions} contains some concluding remarks.  In
the appendix, for mathematical completeness, we prove that the
far-zone flux formulae and the balance equations admit more general
solutions if one relaxes the requirement that the reactive
acceleration be a power series in the individual masses of the binary
or, equivalently, that it be nonlinear in the total mass.

\section{IW method for reactive terms in the equations of motion}
\label{sec:IW}

\subsection{The Procedure}

We consider only two-body systems containing objects that are
sufficiently small that finite-size effects, such as spin-orbit,
spin-spin, or tidal interactions can be ignored.  The dynamics
of such systems is well studied and the two-body equations of motion
conveniently cast into a relative one-body equation of motion is given by :
\begin{equation}
{\bf a} = {\bf a}_N + {\bf a}_{\rm PN}^{(1)} + 
{\bf a}_{\rm 2PN}^{(2)}
+ {\bf a}_{\rm RR}^{(2.5)}
+{\bf a}_{\rm 3PN}^{(3)}
+ {\bf a}_{\rm 1RR}^{(3.5)}
+{\bf a}_{\rm 4PN}^{(4)}+{\bf a}_{\rm Tail}^{(4)}+
+ {\bf a}_{\rm 2RR}^{(4.5)}
 + O(\epsilon^5) \,,
\label{aPNgeneral}
\end{equation}
where the subscripts denote the nature of the term, post-Newtonian
(PN), post-post-Newtonian (2PN), 
Newtonian radiation reaction (RR), post-Newtonian radiation reaction (1RR),
2PN radiation reaction (2RR), tail radiation reaction and so on; and
the superscripts denote the order in $\epsilon$.  
For our purpose we need to know explicitly the
acceleration terms through 2PN order  and they are given by
\cite{dd81,grishchuk,LW90} ($G=c=1$)
\begin{mathletters}
\label{aPN}
\begin{eqnarray}
{\bf a}_N = && - {m \over r^2} {\bf n} \,, \label{aN}
\\
{\bf a}_{\rm PN}^{(1)} = && - {m \over r^2} \biggl\{  {\bf n} \left[
-2(2+\eta)
{m \over r} + (1+3\eta)v^2 - {3 \over 2} \eta \dot r^2 \right]
  -2(2-\eta) \dot r {\bf v} \biggr\} \,, \label{a1PN} \\
{\bf a}_{\rm 2PN}^{(2)} = && - {m \over r^2} \biggl\{ {\bf n} \biggl[ {3
\over 4}
(12+29\eta) \left({m \over r}\right)^2 + \eta(3-4\eta)v^4 + {15 \over 8}
\eta(1-3\eta)
\dot r^4 \nonumber \\
 && - {3 \over 2} \eta(3-4\eta)v^2 \dot r^2 
- {1 \over 2} \eta(13-4\eta) {m \over r} v^2 - (2+25\eta+2\eta^2)
{m \over
r} \dot r^2 \biggr] \nonumber \\
 && - {1 \over 2} \dot r {\bf v} \left[ \eta(15+4\eta)v^2 -
(4+41\eta+8\eta^2)
{m \over r} -3\eta(3+2\eta) \dot r^2 \right] \biggr\} \,, 
\label{a2PN}
\end{eqnarray}
\end{mathletters}
where $\mu\equiv m_1m_2/m$ is
the reduced mass, with $\eta=\mu/m$, and
${\bf n}={\bf x}/r$.  
The $n.5$PN reactive accelerations  are determined by
following the `What else can it be ?' procedure employed in IW which
we summarise here. 
One writes down a general
form for the Newtonian ($\epsilon^{2.5}$), $1$PN 
($\epsilon^{3.5}$) and $2$PN ($\epsilon^{4.5}$)
 radiation-reaction terms in the equations of motion
for two bodies, ignoring tidal and spin effects. 
For the relative
acceleration ${\bf a} \equiv {\bf a}_1- {\bf a}_2$, one assumes the provisional
form
\begin{equation}
{\bf a}=-{8 \over 5} \eta (m/r^2)(m/r)\bigl[-(A_{2.5}+A_{3.5}+ A_{4.5})\dot r
{\bf n} + (B_{2.5}+B_{3.5}+ B_{4.5}){\bf v}\bigr] \,.
\label{provisional}
\end{equation}
The form of Eq.~(\ref{provisional}) is dictated by
the fact that it must be a correction to the Newtonian acceleration
(i.e., be proportional to $m/r^2$), must
vanish in the test body limit when gravitational radiation vanishes
(i.e., be proportional to $\eta$), must be
dissipative, or odd in velocities (i.e., contain the factors
$\dot{r}$, $\bf n$ and  $\bf v$ linearly)  and finally, 
must be related to the
emission of gravitational radiation or be nonlinear
in Newton's constant $G$ (i.e., contain another
factor $m/r$). 
The last condition may be more precisely stated by requiring that
the reactive acceleration be a power series in the individual
masses $m_1$ and $m_2$\cite{tdpvt}.
For spinless, structureless
bodies, the acceleration must lie in the orbital plane (i.e.,
depend only on the vectors $\bf n$ and $\bf v$).
The prefactor $8/5$ is chosen for convenience.  To make the leading
term of $O(\epsilon^{2.5})$ beyond Newtonian order, $A_{2.5}$ and
$B_{2.5}$ must be of $O(\epsilon)$.  For this structureless two-body
system the only variables in the problem
of this order are $v^2$, $m/r$, and $\dot r^2$.  Thus $A_{2.5}$ and
$B_{2.5}$  can each be a linear combination of these three
terms; to those terms we assign six ``Newtonian radiation reaction''
 parameters.  Proceeding similarly, 
 $A_{3.5}$ and $B_{3.5}$ must be of $O(\epsilon^2)$, hence
must each be a linear combination of the six terms $v^4$, $v^2m/r$, 
$v^2\dot r^2$, $\dot r^2m/r$, $\dot r^4$, and $(m/r)^2$.  To these we
assign 12 ``$1$PN RR'' parameters.  
And finally, $A_{4.5}$ and $B_{4.5}$ must be of $O(\epsilon^3)$, 
each  a linear combination of the 10 terms
$v^6$, $v^4\dot r^2$, $v^4m/r$, $v^2\dot r^4$, $v^2(m/r)^2$, 
$v^2\dot r^2(m/r)$, 
$\dot r^6 $, $\dot r^4(m/r)$, $\dot r^2(m/r)^2$ and $(m/r)^3$ to which
we assign $20$ ``$2$PN RR'' parameters. 
The $6$ Newtonian RR and $12$ post-Newtonian  RR parameters were first
determined in IW\cite{iwl,iw95}. This
solution has been checked and reproduced in the preliminary
part of this investigation and constitutes an input
to supplement the conservative acceleration terms in Eq.~(\ref{provisional}) 
for the present study.
Our aim is to evaluate these $20$ parameters appearing in 
$A_{4.5}$ and $B_{4.5}$ that will determine the 2PN radiation
reaction.
It is worth pointing out that in the calculation we are setting up,
the terms in the equations of
motion of $O(\epsilon^3)$ and $O(\epsilon^4)$ 
beyond Newtonian order  do not play any role. The former is non-dissipative
but not yet computed; the latter on the other hand includes dissipative
parts due to the `tail' effects\cite{bd92,aw92,ep93,bs93}
 which have been separately balanced by
the tail luminosity in the works of Blanchet and Damour\cite{bd92,lb96}. 
However all the radiation
reaction results
will remain as `partial results' in the saga of equations of motion 
 until a complete
treatment \`{a} la Chandrasekhar\cite{ce70} and Damour\cite{td82}
 is available through 3PN
order and later through 4PN order.

Through 2PN order, the equations of motion can be derived from a 
generalized Lagrangian that depends not only on  positions
and velocities but also on accelerations. To this order, that is
in the absence of radiation reaction, the Lagrangian leads to
a conserved energy and angular momentum given by
\cite{dd81,grishchuk,KWW}
\begin{mathletters}
\label{EJgeneral}
\begin{eqnarray}
E &=&  E_N +  E_{\rm PN} + E_{\rm 2PN} \,,\\ 
 {\bf J} &=& {\bf J}_N + {\bf J}_{\rm PN} + {\bf J}_{\rm 2PN} \,,
\end{eqnarray}
\end{mathletters}
 where
\begin{mathletters}
\label{EJ}
\begin{eqnarray}
E_N &= & \mu \left ( {1 \over 2} v^2 - {m \over r} \right ) \,,  \label{EN}\\
E_{\rm PN}& = & \mu \left\{
{3 \over 8}
(1-3\eta) v^4
+{1 \over 2} (3+\eta) v^2 {m \over r} +
{1 \over 2} \eta {m \over r} \dot r^2 + {1 \over 2} \left({m \over r}\right)^2
\right\} \,, \label {E1PN}\\
E_{\rm 2PN}& =& \mu \biggl\{ {5 \over 16}(1-7\eta+13\eta^2) v^6
+ {1 \over 8} (21-23\eta-27\eta^2) {m \over r} v^4 \nonumber\\  
&&+ {1 \over 4} \eta (1-15\eta) {m \over r} v^2 \dot r^2
 - {3 \over 8} \eta (1-3\eta){m \over r} \dot r^4  
 - {1 \over 4} (2+15\eta) \left( {m \over r} \right) ^3 \nonumber \\
&&+ {1 \over 8} (14-55\eta+4\eta^2) \left( {m \over r} \right) ^2 v^2  
+ {1 \over 8} (4+69\eta+12\eta^2) \left( {m \over r} \right) ^2 \dot r^2
 \biggr\} \,, \\
 {\bf J}_N& =& {\bf L}_N \,, \label{JN} \\
{\bf J}_{\rm PN} &= & {\bf L}_N \left\{
{1 \over 2} v^2
(1-3\eta) + (3+\eta) {m \over r} \right\} \,, \label{J1PN} \\
{\bf J}_{\rm 2PN}& = & {\bf L}_N \biggl\{
{1 \over 2}
(7-10\eta-9\eta^2) {m \over r}
v^2  - {1 \over 2}\eta (2+5\eta) {m \over r} \dot r^2 \nonumber\\ &&
+ {1 \over 4} (14-41\eta+4\eta^2)
\left( {m \over r} \right)^2
+{3 \over 8} (1-7\eta+13\eta^2) v^4 \biggr\} \,, \label{J2PN}
\end{eqnarray}
\end{mathletters}
and where ${\bf L}_N \equiv \mu {\bf x} \times {\bf v}$.

Through 2PN order, {\em the orbital energy and angular
momentum per unit reduced mass, }  $\tilde E\equiv E/\mu = {1 \over 2}
v^2-m/r+O(\epsilon^2)+O(\epsilon^3)$, ${\bf \tilde J} = {\bf
x}\times{\bf v} [1+O(\epsilon)+O(\epsilon^2)]$,
are constant, and correspond to asymptotically measured quantities.
However, the radiation reaction terms lead to non-vanishing
expressions for
$d\tilde E/dt$ and $d{\bf \tilde J}/dt$ containing the 20 undetermined
parameters.
Following IW, starting from the 
2PN-conserved expressions for $\tilde E$ and ${\bf\tilde J}$ we
calculate $d\tilde E/dt$ and $d{\bf \tilde J}/dt$ using the
2PN two-body equations of motion \cite{dd81,grishchuk,LW90}
supplemented by the radiation-reaction terms of Eq.~(\ref{provisional}).
In the balance approach, this time variation of the `conserved' quantities
is equated to the negative of the flux of energy and angular momentum 
carried by the gravitational waves to the far-zone. 
Thus in addition to the EOM and conserved quantities we need the
2PN accurate expressions for the far-zone fluxes of energy and angular
momentum for a system of two particles moving on 
general quasi--elliptic orbits. The waveform, energy and angular momentum
flux   have been computed
by Gopakumar and Iyer\cite{gi96} using the BDI\cite{bdigen,bdi2pn} formalism,
and independently the waveform and energy flux 
by Will and Wiseman\cite{ww2pn} using their new improved version of
the EW\cite{ew75} formalism. We quote below the final results 
for the {\em fluxes per unit reduced mass}:
\begin{mathletters}
\label{flux-schem}
\begin{eqnarray}
\biggl (\frac{d{\cal E }}{dt}\biggr )_{\rm far-zone} &=& \dot{{\cal E}}_N 
+ \dot{{\cal E}}_{\rm 1PN}+
\dot{{\cal E}}_{\rm 1.5PN}
 + \dot{{\cal E}}_{\rm 2PN} \,,\\ [1em]
\biggl (\frac{d{\bf {\cal J}}}{dt}\biggr )_{\rm far-zone} &=& 
{\bf \tilde L}_N\left[\dot{{ \cal J}}_N +
\dot{{ \cal J}}_{\rm 1PN}+ 
\dot{{ \cal J}}_{\rm 1.5PN} + \dot{{ \cal J} }_{\rm 2PN}\right] \,,
\end{eqnarray}
\end{mathletters}
where
\begin{mathletters}
\label{2PNfluxes}
\begin{eqnarray}
\dot{{\cal E}}_N &=& {8\over5}\eta{{m^2}\over r^3}{m\over r} 
  \biggl( 4v^2 - {11\over 3}\dot r^2 \biggr) \,, \\
\dot{{\cal E}}_{\rm 1PN} &=& {8\over5}\eta{{m^2}\over r^3}{m\over r}
  \biggl[\frac{1}{84}(785-852\eta)v^4 
    - \frac{1}{42}(1487-1392\eta)v^2\dot r^2
    -\frac{40}{21}(17-\eta)v^2\frac{m}{r}\nonumber\\
&&  +\frac{1}{28}(687-620\eta)\dot r^4
    +\frac{2}{21}(367-15\eta)\dot r^2\frac{m}{r}
    +\frac{4}{21}(1-4\eta)\left(\frac{m}{r}\right)^2
  \biggr]\,, \\
\dot{{\cal E}}_{\rm 2PN} &=& {8\over5}\eta{{m^2}\over r^3}{m\over r}
  \biggl[\frac{1}{126}(1692-5497\eta+4430\eta^2)v^6
     -\frac{1}{42}(1719-10278\eta+6292\eta^2)v^4\dot{r}^2 \nonumber\\
&&   -\frac{1}{63}(4446-5237\eta+1393\eta^2)v^4\frac{m}{r}
     +\frac{1}{42}(2018-15207\eta+7572\eta^2)v^2\dot{r}^4 \nonumber\\
&&   +\frac{1}{21}(4987-8513\eta+2165\eta^2)v^2\dot{r}^2\frac{m}{r}
     +\frac{1}{2268}(281473+81828\eta+4368\eta^2)v^2
        \left(\frac{m}{r}\right)^2 \nonumber\\
&&   -\frac{1}{126}(2501-20234\eta+8404\eta^2)\dot{r}^6
     -\frac{1}{189}(33510-60971\eta+14290\eta^2)\dot{r}^4
        \frac{m}{r}\nonumber\\
&&   -\frac{1}{756}(106319+9798\eta+5376\eta^2)\dot{r}^2
        \left(\frac{m}{r}\right)^2 \nonumber \\
&&   -\frac{2}{189}(253-1026\eta+56\eta^2)
        \left(\frac{m}{r}\right)^3
  \biggr] \,,\\
\dot{{\cal J}}_N &=& \frac{8}{5}\eta\frac{m}{r^2}\frac{m}{r}
  \biggl(2v^2-3\dot{r}^2+2\frac{m}{r}\biggr)\,,\\
\dot{{\cal J}}_{\rm 1PN} &=& \frac{8}{5}\eta\frac{m}{r^2}\frac{m}{r}
  \biggl[\frac{1}{84}(307-548\eta)v^4
     -\frac{1}{14}(74-277\eta)v^2\dot{r}^2 
     -\frac{1}{21}(58+95\eta)v^2\frac{m}{r} \nonumber\\
&&   +\frac{1}{28}(95-360\eta)\dot{r}^4 
     +\frac{1}{42}(372+197\eta)\dot{r}^2\frac{m}{r}
     -\frac{1}{42}(745-2\eta)\left(\frac{m}{r}\right)^2
  \biggr] \,,\\
\dot{{\cal J}}_{\rm 2PN} &=& \frac{8}{5}\eta\frac{m}{r^2}\frac{m}{r}
  \biggl[\frac{1}{504}(2665-12355\eta+12894\eta^2)v^6
     -\frac{1}{168}(2246-12653\eta+15637\eta^2)v^4\dot{r}^2\nonumber\\
&&   +\frac{1}{504}(165-491\eta+4022\eta^2)v^4\frac{m}{r}
     +\frac{1}{168}(3575-16805\eta+15680\eta^2)v^2\dot{r}^4\nonumber\\
&&   +\frac{1}{504}(21853-21603\eta+2551\eta^2)v^2\dot{r}^2\frac{m}{r}
     -\frac{1}{252}(10651-10179\eta+3428\eta^2)v^2
        \left(\frac{m}{r}\right)^2 \nonumber\\
&&   -\frac{5}{18}(39-163\eta+97\eta^2)\dot{r}^6
     -\frac{1}{504}(22312-41398\eta+9695\eta^2)\dot{r}^4
        \frac{m}{r} \nonumber\\
&&   +\frac{1}{252}(8436-25102\eta+4587\eta^2)\dot{r}^2
        \left(\frac{m}{r}\right)^2 \nonumber \\
&&   +\frac{1}{2268}(170362+70461\eta+1386\eta^2)
        \left(\frac{m}{r}\right)^3
  \biggr] \,.
\end{eqnarray}
\end{mathletters}
In the above expressions
${\bf \tilde L}_N={\bf L}_N /\mu$ and the tail terms are not listed.
It is important to emphasize that the `tail' contribution to the 
reaction force 
is such that the balance equation for energy is verified for the 
tail luminosity\cite{bd92,lb96}. This corresponds to the `tail'
acceleration at $4$PN. With this part independently accounted for,
in our analysis  we focus
on the `instantaneous' terms without loss of generality.
It is worth recalling that the `balance' one sets up in the above treatment
is always modulo total time derivatives of the variables involved. 
This is crucial to realize and in IW this was systematically
accounted for by noting that 
at orders of approximation beyond those at which
they are strictly conserved (and thus well defined), $\tilde E$ and
${\bf \tilde J}$ are ambiguous upto such terms.  
Consequently, we have the freedom
to add to $\tilde E$ and ${\bf \tilde J}$ arbitrary terms of order
$\epsilon^{2.5}$,  $\epsilon^{3.5}$, and $\epsilon^{4.5}$ 
beyond the Newtonian expressions
without affecting their conservation at 2PN order.
There are three such terms of the appropriate general form
at $O(\epsilon^{2.5})$ in each of $\tilde E$ and
${\bf \tilde J}$, respectively, 6 each at $O(\epsilon^{3.5})$
and 10 each at $O(\epsilon^{4.5})$, resulting in 6 
additional Newtonian RR parameters, 12 additional $1$PN RR parameters
and $20$  additional $2$PN RR parameters, respectively. As discussed in
detail in the following section, these numbers are very much tied up
with the `functional form' we assume for the ambiguous terms and in this
section we follow IW in close detail.
Equating time derivatives of the 
resulting generalized energy and angular
momentum expressions 
$\tilde E^*$ and
${\bf \tilde J^*}$
( rather than only the conserved expressions)
to the negative
of the  far-zone flux formulae
and comparing them term by
term one seeks to determine  the extent to which one can 
deduce the 4.5PN reactive acceleration terms by the refined
balance approach.

\subsection{The 2PN RR computation and results}

The above procedure is implemented order by order. 
All the computations were done with MAPLE \cite {Map} and independently
checked by MATHEMATICA \cite {Math}. 
At the leading
order, when the flux is given by the  quadrupole equation, one deduces
the `Newtonian RR'  or 2.5PN term in the acceleration. 
In this case, in addition to the 6 unknowns in the reactive acceleration,
one has 3 unknowns each for the possible 2.5PN ambiguities  in the
$\tilde E^*$ and $\bf \tilde J^*$.
As demonstrated in IW, the balance equations yield
12 constraints on these 12 Newtonian RR
parameters. Of the 12 constraints, only 10 are linearly independent,
and thus finally one obtains 10 linear inhomogeneous equations 
for 12 Newtonian radiation reaction variables.
Solving these equations one obtains explicit forms for
$A_{2.5}$, 
$B_{2.5}$ and  $\tilde E_{2.5}$, $\tilde J_{2.5}$
in terms of two $2.5$PN arbitrary parameters.
To get the $3.5$PN reactive terms, one adopts the above solution and extends
the calculation to $O(\epsilon^{3.5})$ after introducing 
$\tilde E_{3.5}$ and $\tilde J_{3.5}$ with  12 additional $1$PN RR parameters.
At 3.5PN there are 20 constraints on the 24 post-Newtonian
radiation reaction parameters; of the 20 only 18
are linearly independent; the solution to this system yields 
explicit forms for $A_{3.5}$,
$B_{3.5}$ and  $\tilde E_{3.5}$, $\tilde J_{3.5}$
in terms of six $3.5$PN arbitrary parameters.
Since we need these results for the present computation,  we reproduce 
them from IW\cite{newnotn}.
\begin{mathletters}
\label{ABexp}
\begin{eqnarray}
A_{2.5} = && 3(1+\alpha_3)v^2+
       {1 \over 3}(23+6\beta_2-9\alpha_3){m\over r}-5\alpha_3\dot r^2\,, 
\label{AN}\\
B_{2.5} = && (2+\beta_2)v^2+(2-\beta_2){m \over r}
-3(1+\beta_2)\dot r^2\,, \label{BN}\\
A_{3.5} = &&f_1v^4+f_2v^2{m \over r}+f_3v^2 \dot r^2+f_4 \dot r^2 {m \over
r}+f_5 \dot r^4+f_6\left({m \over r}\right)^2 \,, \label{APN}\\
B_{3.5} = &&g_1v^4+g_2v^2{m \over r}+g_3v^2 \dot r^2+g_4 \dot r^2 {m \over
r}+g_5 \dot r^4+g_6\left({m \over r}\right)^2 \,, \label{BPN} 
\end{eqnarray}
\end{mathletters}
where
\begin{mathletters}
\label{cdsolution}
\begin{eqnarray}
f_1=&&{1 \over 28} (117+132\eta)-{3 \over 2} \alpha_3 (1-3\eta)+
3\xi_2-3\rho_5\,,
\\
f_2=&&-{1 \over 42} (297-310\eta)-3\beta_2(1-4\eta)-{3 \over 2} \alpha_3(7+13\eta)
-2\xi_1-3\xi_2+3\xi_5+3\rho_5\,,
\\
f_3=&&{5 \over 28}(19-72\eta)+{5 \over 2}
\alpha_3(1-3\eta)-5\xi_2+5\xi_4+5\rho_5\,,
\\
f_4=&&-{1 \over 28}(687-368\eta)-6\beta_2\eta+{1
\over2}\alpha_3(54+17\eta)-2\xi_2-5\xi_4-6\xi_5\,,
\\
f_5=&&-7\xi_4\,,
\\
f_6=&&-{1 \over 21}(1533+498\eta)-\beta_2(14+9\eta)+3\alpha_3(7+4\eta)
-2\xi_3-3\xi_5\,,
\\
g_1=&&-3(1-3\eta)-{3 \over 2}\beta_2(1-3\eta)-\xi_1\,,
\\
g_2=&&-{1 \over 84}(139+768\eta)-{1 \over 2}\beta_2(5+17\eta)
+\xi_1-\xi_3\,,
\\
g_3=&&{1 \over 28}(369-624\eta)+{3 \over 2}(3\beta_2+2\alpha_3)(1-3\eta)
+3\xi_1-3\rho_5\,,
\\
g_4=&&{1 \over 42}(295-335\eta)+{1
\over2}\beta_2(38-11\eta)-3\alpha_3(1-3\eta)
+2\xi_1+4\xi_3+3\rho_5\,,
\\
g_5=&&{5 \over 28}(19-72\eta)-5\alpha_3(1-3\eta)+5\rho_5\,,
\\
g_6=&&-{1 \over 21}(634-66\eta)+\beta_2(7+3\eta)+\xi_3\,.
\end{eqnarray}
\end{mathletters}
The quantities $\alpha_3$, $\beta_2$, $\xi_1$, $\xi_2$, $\xi_3$, $\xi_4$,
$\xi_5$ and $\rho_5$ are parameters that represent the unconstrained
degrees of freedom that correspond to gauge transformations.
In addition to the reactive terms listed above, 
one of the coefficients that determine the 2.5PN ambiguity 
in $\tilde E$ and $\bf\tilde J$
and three of the coefficients
that determine the corresponding 3.5PN ambiguity  
are nonvanishing. We list these also since they are needed for setting
up the 4.5PN computation:
\begin{mathletters}
\label{3.5amb}
\begin{eqnarray}
\alpha_1 = && -\left( 2 +\beta_2\right)\,,\\
\xi_{6}=&& -{4\over 21}\left( 1 -4\eta\right)\,,\\
 \rho_3 =&& \xi_1+ {1\over 84}\left( 307 -548\eta\right)\,,\\ 
 \rho_6 =&& \xi_3 -{1 \over 42}( 271 -214\eta)\,.
\end{eqnarray}
\end{mathletters}

We now adopt the 2.5PN and 3.5PN solutions  given by 
Eqs.~(\ref{ABexp}), (\ref{cdsolution}) and (\ref{3.5amb}). Following the
IW strategy, we
assume the 4.5PN terms in the equations of motion to be of the form
\begin{mathletters}
\label{AB2PN}
\begin{eqnarray}  
A_{4.5} &=& h_1v^6 + h_2v^4\dot r^2 + h_3v^4{m\over r} + 
            h_4v^2\dot r^4 + h_5v^2\left({m\over r}\right)^2 \nonumber\\
        &&\quad  + h_6v^2\dot r^2{m\over r} +
            h_7\dot r^6 + h_8\dot r^4{m\over r} 
            + h_9\dot r^2\left({m\over r}\right)^2 +
  h_{10}\left({m\over r}\right)^3  \,, \label{A2PN}\\
B_{4.5} &=&k_1v^6 + k_2v^4\dot r^2 + k_3v^4{m\over r} + 
  k_4v^2\dot r^4 + k_5v^2\left({m\over r}\right)^2 \nonumber \\
&&\quad + k_6v^2\dot r^2{m\over r} +
  k_7\dot r^6 + k_8\dot r^4{m\over r} + k_9\dot r^2\left({m\over r}\right)^2 +
  k_{10}\left({m\over r}\right)^3  \,. \label{B2PN}
\end{eqnarray}
\end{mathletters}
We also assume for the ambiguity in
$\tilde E_{4.5}$ and $\tilde J_{4.5}$ the restrictions and 
functional forms adopted in IW
and also require that $\bf \tilde J$ remain a pseudo-vector. 
The  `generalized' ``energy'' and ``angular momentum'' through
4.5PN are thus given as sums of the conserved
parts Eqs. (~\ref{EJ}), the `most general' 2.5PN and 3.5PN contributions --
i.e., with coefficients  determined by the Newtonian RR 
and 1PN RR calculations,
and arbitrary 4.5PN terms. 
We use $\tilde E^*$ and $\bf \tilde J^*$ to distinguish these
quantities from the conserved energy and angular momentum.
We get (per unit reduced mass)
\begin{mathletters}
\label{EJ*4.5}
\begin{eqnarray}
\tilde E^* \equiv &&\tilde E_N +\tilde E_{\rm PN} + \tilde E_{\rm 2PN} + 
\tilde E_{2.5} + \tilde E_{3.5} + \tilde E_{4.5} \nonumber \\
= &&\tilde E_N +\tilde E_{\rm PN} + \tilde E_{\rm 2PN}
+{8 \over 5} \eta \left({m \over r}\right)^2 
\dot r  [ (2+\beta_2) v^2 -\alpha_3 \dot r^2 ] \nonumber \\
&&-{8 \over 5} \eta \left({m \over r}\right)^2 \dot r 
\biggl [ \xi_1v^4+\xi_2v^2
\dot r^2+\xi_3v^2{m \over r}+\xi_4 \dot r^4+\xi_5 \dot r^2 {m\over r}
-{4\over21}(1-4\eta)\left({m \over r}\right)^2 \biggr ] \nonumber \\
&&- {8 \over 5} \eta \left({m \over r}\right)^2 \dot r 
\biggl[
  \psi_1v^6 + \psi_2v^4\dot r^2 + \psi_3v^4{m\over r} + 
  \psi_4v^2\dot r^4 + \psi_5v^2\left({m\over r}\right)^2 
  + \psi_6v^2\dot r^2{m\over r} +
  \psi_7\dot r^6\nonumber \\
&&\qquad +\psi_8\dot r^4{m\over r} + \psi_9\dot r^2\left({m\over r}\right)^2 +
  \psi_{10}\left({m\over r}\right)^3 
\biggr] \,, \label{E*9/2}\\
{\bf \tilde J^*} \equiv &&{\bf \tilde J}_N + {\bf \tilde J}_{\rm PN} +
{\bf \tilde J}_{\rm 2PN} + {\bf\tilde J}_{2.5}+ {\bf \tilde J}_{3.5} + 
{\bf \tilde J}_{4.5} \nonumber \\
=&&{\bf \tilde J}_N +{\bf \tilde J}_{\rm PN} + {\bf \tilde J}_{\rm 2PN}
+{8 \over 5} \eta {\bf \tilde L}_N  {m \over r}  \dot r 
\left(\beta_2 {m \over r}\right) \nonumber \\
&&-{8 \over 5} \eta {\bf \tilde L}_N  {m \over r}  \dot r 
\biggl[ {1\over84}(307-548\eta+84\xi_1)v^2{m \over r}+
\rho_5 \dot r^2 {m\over r}-
{1\over42}(271-214\eta-42\xi_3)
\left({m \over r}\right)^2 \biggr]\nonumber \\ 
&&-{8 \over 5} \eta {\bf \tilde L}_N  {m \over r}  \dot r
\biggl[
  \chi_1v^6 + \chi_2v^4\dot r^2 + \chi_3v^4{m\over r} + 
  \chi_4v^2\dot r^4 + \chi_5v^2\left({m\over r}\right)^2 + 
  \chi_6v^2\dot r^2{m\over r} +
  \chi_7\dot r^6\nonumber \\
&&\qquad +\chi_8\dot r^4{m\over r} + \chi_9\dot r^2\left({m\over r}\right)^2 +
  \chi_{10}\left({m\over r}\right)^3 \biggr] \,, \label{J*9/2}
\end{eqnarray}
\end{mathletters}
      We now compute the 4.5PN terms in $d\tilde E^*/dt$
and $d{\bf\tilde J^*}/dt$
using the identities 
\begin{mathletters}
\label{ident}
  \begin{eqnarray}
    {1 \over 2}{dv^2\over dt} \equiv&& {\bf v} \cdot {\bf a}\,,\\
    {d({\bf x} \times {\bf v})\over dt}\equiv&&{\bf x} \times 
                       {\bf a}\,,\\
    \ddot r \equiv&& {v^2 + {\bf r}\cdot{\bf a} - \dot r^2 \over r}\,,
  \end{eqnarray}
\end{mathletters}
where $ \bf a $  is given by Eqs.~(\ref{aPNgeneral}), (\ref{aPN}), 
(\ref{provisional}), (\ref{ABexp}), (\ref{cdsolution}) and (\ref{AB2PN}). 
To compute  $\dot{\tilde E}{}^*$ and $\dot{\bf\tilde J}{}^*$ to
$O(\epsilon^{4.5})$,  one needs to evaluate
$(\dot{\tilde E}{}_{N}$,$\dot{{\bf\tilde J}}{}_{N})$,
$(\dot{\tilde E}{}_{\rm 1PN}$, $\dot{{\bf\tilde J}}{}_{\rm 1PN})$ and 
$(\dot{\tilde E}{}_{\rm 2PN}$, $\dot{{\bf\tilde J}}{}_{\rm 2PN})$ by using
{\bf a} to $O(\epsilon^{4.5})$,
$O(\epsilon^{3.5})$ and
$O(\epsilon^{2.5})$, respectively. 
On the other hand, for time derivatives of the `ambiguity parts', 
$ (\tilde E_{4.5} $, ${\bf {\tilde J}} _{4.5}) $, 
$ (\tilde E_{3.5} $, ${\bf {\tilde J}} _{3.5}) $
and $ (\tilde E_{2.5} $, ${\bf {\tilde J}} _{2.5}) $,
the relevant
accelerations are the `conservative' accelerations to
order Newtonian, post-Newtonian and second post-Newtonian,
respectively.
Schematically, we get,
\begin{mathletters}
\label{EJ*dot9/2}
\begin{eqnarray}
{{d \tilde E^*} \over {dt}}&=& -{8 \over 15} \eta {{m^2} \over r^3}
\biggl [
{m \over r}\biggl ( 12v^2 -11 \dot r^2 \biggr )
+ {m \over r}\left\{\frac{1}{28}\left[(785 -
852\eta)v^4
+2(-1487 + 1392\eta)v^2\dot{r}^2 \right.\right.\nonumber\\
& &\left.\left. + 160(-17 +\eta) {m \over r}v^2
+ 3(687 - 620\eta)\dot{r}^4 + 8 (367 - 15\eta){m \over r}\dot r^2
\right.\right.\nonumber\\
& & \left.\left. +16(1 - 4\eta)\left({m \over r}\right)^2\right] \right\}
+\sum_{i=1}^{15}{\cal R}^{[4.5]}_i  {\cal Y}^{[4]}_i
\biggr ]\,, \label{E*dot9/2} \\
{{d {\bf \tilde J^*}} \over {dt}}&=& -{8 \over 5} \eta 
{\bf \tilde L}_N
{{m} \over r^2} \biggl [
{m \over r}\biggl ( 2v^2 +2{m \over r} -3 \dot r^2  \biggr )+
{m \over r}\left\{ {1 \over 84}
\left[ (307 - 548\eta)v^4
+ 6\left(-74 + 277\eta \right)v^2\dot{r}^2 \right.\right.\nonumber\\
&& - 4(58 + 95\eta){m \over r}v^2 
+ 3(95 - 360\eta)\dot{r}^4 + 2(372 + 197\eta)
{m \over r}\dot{r}^2 \nonumber \\
&& \left.\left.
+ 2(-745 + 2\eta)\left({m \over r}\right)^2 \right]\right\}
+\sum_{i=1}^{15}{\cal S}^{[4.5]}_i  {\cal Y}^{[4]}_i
 \biggr ]\,, \label{J*dot9/2}
\end{eqnarray}
\end{mathletters}
 where  
\begin{eqnarray}
\label{cy4}
 {\cal Y}^{[4]}_i (i=1\ldots15)= 
\biggl[&&v^8, v^6\left({m \over r}\right), v^6\dot r^2,
v^4\left({m \over r}\right)^2,
v^4\dot r^4, v^4\left({m \over r}\right)\dot r^2,
v^2 \left({m\over r}\right)^3, v^2\dot r^6,\nonumber\\
&&v^2\left({m \over r}\right)^2\dot r^2,
v^2\left({m \over r}\right)\dot r^4,
\left({m \over r}\right)^4,
\left({m \over r}\right)^3\dot r^2,\left({m \over r}\right)^2\dot r^4,
\left({m \over r}\right)\dot r^6,\dot r^8 \biggr ]\,,
\end{eqnarray}
and ${\cal R}^{[4.5]}_i$ and ${\cal S}^{[4.5]}_i$
consist of combinations of the parameters
$h_i$ and $k_i$ from $A_{4.5}$ and $B_{4.5}$, 
$\psi_i$, $\chi_i$ combined with functions of $\eta$ from 
$\tilde E_{4.5}$ and ${\bf \tilde J}_{4.5}$,
$\xi_1,\xi_2,\xi_3,\xi_4,\xi_5,\rho_5$ combined with
functions of $\eta$ from 1PN corrections
of 3.5PN terms and 
$\alpha_3$ and $\beta_2$
combined with functions of $\eta$ from 2PN corrections of 2.5PN terms.
We equate $d\tilde E^*/dt$ and $d{\bf\tilde J^*}/dt$
thus obtained to the negative of the 
$2$PN far-zone fluxes given by Eqs.~(\ref{2PNfluxes}).
This results in 30 constraints on the 40 parameters $h_i$, $k_i$,
$\psi_i$ and $\chi_i$. Two of these constraints being redundant,
of the 30 constraints only 28 are
linearly independent. The system of 28 linear inhomogeneous equations for
40 variables is therefore under-determined to the extent of 12 arbitrary
parameters, and we choose these to be $\psi_1\ldots\psi_9$, $\chi_6$,
$\chi_8$ and $\chi_9$. With this choice, the coefficients in
Eq. (~\ref{AB2PN} determining the $4.5$PN reactive acceleration are given by
\begin{mathletters}
\label{sol4.5}
\begin{eqnarray}
h_1 = && -{1\over168}(121-2278\eta+4012\eta^2)
         -{3\over8}\alpha_3(1-9\eta+21\eta^2)
         -{3\over2}(\xi_2-\rho_5)(1-3\eta)\nonumber\\
      && \quad + 3\psi_2 - 3\chi_6\,,\\
h_2 = &&  {5\over84}(329-1487\eta+1244\eta^2)
         +{5\over8}\alpha_3(1-9\eta+21\eta^2)
         +{5\over2}(\xi_2-\xi_4-\rho_5)(1-3\eta)\nonumber\\
      && \quad -5\psi_2 + 5\psi_4 
         + 5\chi_6 - 5\chi_8\,,\\
h_3 = &&  {1\over504}(7692-87429\eta+11218\eta^2)
         +{3\over8}\alpha_3(1-97\eta+25\eta^2)
         +{1\over4}\beta_2(3-3\eta-19\eta^2)\nonumber\\
      && \quad+3\xi_1(1-4\eta)
         -{3\over2}(\xi_2-\rho_5)(7+13\eta)
         -{3\over2}\xi_5(1-3\eta)\nonumber \\
      && \quad - 2\psi_1 - 3\psi_2 + 3\psi_6
         +3\chi_6 - 3\chi_9\,,\\
h_4 = && -{5\over18}(39-163\eta+97\eta^2)
         +{7\over2}\xi_4(1-3\eta) - 7\psi_4 + 7\psi_7 + 7\chi_8\,,\\
h_5 = && -{1\over252}(37089-64005\eta+11297\eta^2)
         +9\alpha_3(2+13\eta+2\eta^2) 
         +{1\over4}\beta_2(48-121\eta-54\eta^2)\nonumber\\
      && \quad+\xi_1(14+9\eta) 
         + 3(\xi_2-\rho_5)(7+4\eta)
         +3\xi_3(1-4\eta) - {3\over2}\xi_5(7+13\eta)\nonumber\\
      && \quad -2\psi_3 - 3\psi_6 + 3\psi_9 + 3\chi_9\,,\\
h_6 = && -{1\over504}(45475-219535\eta+43121\eta^2)
         -{1\over4}\alpha_3(14-403\eta+77\eta^2)
         -{3\over2}\beta_2\eta(7-13\eta)\nonumber\\
      && \quad  + 6\eta\xi_1
         +{1\over2}\xi_2(68-9\eta) 
         - {5\over2}\xi_4(7+13\eta)
         +3\xi_5(1-3\eta) - {1\over2}\rho_5(62+19\eta)\nonumber\\
      && \quad -4\psi_2 - 5\psi_4 - 6\psi_6 + 5\psi_8 + 2\chi_6 
         +5\chi_8 +6\chi_9\,,\\
h_7 = && -9\psi_7\,,\\
h_8 = &&  {1\over252}(5002-36589\eta+4496\eta^2)
         -{1\over8}\alpha_3\eta(233-63\eta)
         +{33\over4}\beta_2\eta(1-3\eta)
         +3\eta\xi_2\nonumber\\ 
      && \quad+{1\over2}\xi_4(82+23\eta)
         + 5\eta\rho_5 - 2\psi_4 - 7\psi_7 - 8\psi_8\,,\\
h_9 = &&  {1\over756}(181371-342479\eta+42598\eta^2)
         -{1\over2}\alpha_3(117+109\eta+6\eta^2)\nonumber\\
      && \quad -{1\over4}\beta_2(28+245\eta+20\eta^2)
         +2\eta\xi_1 
         + (2\xi_2+5\xi_4)(7+4\eta) + 7\eta\xi_3
         +{1\over2}\xi_5(60+21\eta) \nonumber\\
      && \quad+ 3\eta\rho_5
         - 2\psi_6
         -5\psi_8 - 7\psi_9\,,\\
h_{10}=&& {1\over756}(265265+262230\eta+15072\eta^2)
         -{3\over4}\alpha_3(102+177\eta+16\eta^2)\nonumber\\
       && \quad +{1\over4}\beta_2(200+325\eta+40\eta^2)
          +\xi_3(14+9\eta) + 3\xi_5(7+4\eta) - 2\psi_5 - 3\psi_9\,,\\
k_1 = &&  {3\over8}(\beta_2+2)(1-\eta-11\eta^2)
         +{3\over2}\xi_1(1-3\eta) - \psi_1\,,\\
k_2 = && -{1\over168}(499-2656\eta-146\eta^2)
         -{3\over2}\alpha_3(1-3\eta-3\eta^2)
         -{9\over8}\beta_2(1-\eta-11\eta^2)\nonumber\\
      && \quad -{3\over2}(3\xi_1-2\xi_2-\rho_5)(1-3\eta) 
         + 3\psi_1 - 3\chi_6\,,\\
k_3 = &&  {1\over504}(81-9127\eta-14482\eta^2)
         -{1\over8}\beta_2(3+121\eta+7\eta^2)
         +{1\over2}\xi_1(5+17\eta)\nonumber\\
      && \quad +{3\over2}\xi_3(1-3\eta) + \psi_1 - \psi_3\,,\\
k_4 = &&  {5\over84}(329-1487\eta+1244\eta^2)
         +{5\over2}\alpha_3(1-3\eta-3\eta^2)
         -{5\over2}(2\xi_2-2\xi_4+\rho_5)(1-3\eta)\nonumber \\
      && \quad + 5\chi_6 - 5\chi_8\,,\\
k_5 = && -{11\over252}(1107-805\eta-508\eta^2)
         +{1\over4}\beta_2(16+255\eta+22\eta^2)
         -\xi_1(7+3\eta) \nonumber \\
      && \quad + {1\over2}\xi_3(5+17\eta)
         + \psi_3 - \psi_5\,,\\
k_6 = &&  {1\over504}(1797+54816\eta-22463\eta^2)
         +{3\over2}\alpha_3(1+3\eta+5\eta^2)
         -{1\over4}\beta_2(42-485\eta+173\eta^2) \nonumber \\
      && \quad -{1\over2}\xi_1(56-49\eta) - 3(\xi_2+2\xi_3-\xi_5)(1-3\eta)
         +{3\over2}\rho_5(7+11\eta) \nonumber\\
      && \quad + 4\psi_1 + 4\psi_3
         +3\chi_6 - 3\chi_9\,,\\
k_7 = && -{5\over18}(39-163\eta+97\eta^2)
         -7\xi_4(1-3\eta) + 7\chi_8\,,\\ 
k_8 = && -{1\over504}(39808-92788\eta+24563\eta^2)
         +{1\over2}\alpha_3(14-105\eta+59\eta^2)
         -{3\over8}\beta_2\eta(69+13\eta)\nonumber \\
      && \quad -3\eta\xi_1
         -(2\xi_2+5\xi_4+6\xi_5)(1-3\eta)
         -{1\over2}\rho_5(62+3\eta) 
         + 2\chi_6 + 5\chi_8 + 6\chi_9\,,\\
k_9 = &&  {1\over252}(8319-7683\eta+11809\eta^2)
         +3\alpha_3(3-13\eta-\eta^2)
         -{1\over4}\beta_2(194+215\eta+24\eta^2)\nonumber\\
      && \quad -(2\xi_1+3\rho_5)(7+3\eta)
         -{1\over2}\xi_3(44-9\eta) - 3\xi_5(1-3\eta)
         +2\psi_3 + 5\psi_5 + 3\chi_9\,,\\
k_{10}=&& {1\over2268}(425413+111636\eta-6912\eta^2)
         -{1\over2}\beta_2(53+103\eta+4\eta^2)\nonumber\\
       && \quad -\xi_3(7+3\eta) + \psi_5\,.
\end{eqnarray}
\end{mathletters}
At the $4.5$PN order,
4 parameters determining $\tilde {E}_{4.5}$
and ${\bf\tilde J}_{4.5}$ are non-vanishing and are given by
\begin{eqnarray}
\label{4.5nrs}
\psi_{10}&=& {1 \over 189}\left( 362 -1548\eta +400\eta^2\right)\,,\nonumber\\
\chi_3 &=& \psi_1 +{1 \over 504}\left( 2665 -12355\eta 
+12894\eta^2\right)\,,\nonumber \\
\chi_5 &=& \psi_3 +{7\over 2}\beta_2\eta -{1 \over 126}\left( 524 
- 4483 \eta +3675\eta^2\right)\,,\nonumber\\
\chi_{10}&=& \psi_5 -{7\over 2}\beta_2\eta +{1 \over 252}\left( 775 -
3939\eta +2942 \eta^2\right)\,. 
\end{eqnarray}
A final minor remark is with regard  to the two possible ways one may
implement the requirement that  the ambiguity in 
$\bf \tilde J^*$ be a pseudovector.
One may either  choose it proportional to ${\bf \tilde L}_N$ as in the
treatment above
or to the conserved angular momentum ${\bf \tilde J}$.
At 2.5PN order both choices are identical. At  the
3.5PN  order, the two choices  lead to an identical system of linear
equations barring a
translation in
the values of $\rho_3 $ and $\rho_6$ 
by an amount given by the coefficients of $v^2$ and $m/r$ in 
$ {\bf J}_{1PN} $:
\begin{eqnarray}
\rho_3 {\rightarrow} \bar {\rho_3}&=& \rho_3 +
{1 \over 2}\left( 1 -3\eta\right)\beta_2 \,,\nonumber \\
\rho_6 {\rightarrow} \bar{\rho_6}&=& \rho_6 +
\left( 3 +\eta\right)\beta_2\,.
\end{eqnarray}
Since $\rho_3 $ and $\rho_6$  are {\it not} among the arbitrary
parameters determining the solution,
the  solution determining the reactive terms
 and $\xi_6$ is {\it unchanged}. Only the expressions for $\rho_3 $ and 
$\rho_6$ are changed to
\begin{eqnarray}
\bar {\rho_3} &=& \xi_1+ {1\over 84}\left( 307 -548\eta\right)
+{1 \over 2}\left( 1 -3\eta\right)\beta_2 \,,\nonumber \\
\bar {\rho_6} &=& \xi_3 -{1 \over 42}( 271 -214\eta )
+ \left( 3 +\eta\right)\beta_2 \,.
\end{eqnarray}
At $4.5$PN  order, however, the situation is different. 
Indeed, as before, the two choices  lead to an identical system of linear
equations barring a
translation in
the values of  the five  parameters $\chi_3 $, $\chi_5$, $\chi_6$, $\chi_9$
and $\chi_{10}$.
\begin{eqnarray}
\chi_3 {\rightarrow} \bar{\chi_3}&=& \chi_3 +{1 \over 8}\left( 1 -9\eta 
+21\eta^2\right)\beta_2 -{1 \over 2}\left(1 -3\eta\right)\xi_1 
-{1 \over 168}\left(307 -1469\eta +1644\eta^2\right)\,,\nonumber \\
\chi_5 {\rightarrow} \bar{\chi_5}&=&\chi_5 +
{1 \over 2}\left(1 +6\eta -3\eta^2\right)\beta_2 -\left(3 +\eta\right)\xi_1
-{1 \over 2}\left(1 -3\eta\right)\xi_3 \nonumber\\
&& \quad -{1 \over 42}\left( 325 -155\eta -595\eta^2\right)\,,\nonumber\\
\chi_6 {\rightarrow} \bar{\chi_6}&=&\chi_6 
-{1 \over 2}\left(1 -3\eta\right)\rho_5 \,,\nonumber\\
\chi_9 {\rightarrow} \bar{\chi_9}&=&\chi_9 
-{1\over 2}\left(2 +5\eta\right)\eta \beta_2 
- \left( 3 +\eta\right)\rho_5 \,,\nonumber\\
\chi_{10} {\rightarrow} \bar{\chi_{10}}&=& \chi_{10}
-{1 \over 4}\left(22 +65\eta\right)\beta_2 -\left(3 +\eta\right)\xi_3
+{1 \over 294}\left( 5691 -2597\eta -1498\eta^2\right) \,.
\end{eqnarray}
Consequently, in terms of the above `shifted' variables, the solutions
for the reactive accelerations are identical.
As  $\chi_6$ and $\chi_9$ 
are among the {\it independent} parameters that
determine the reactive acceleration, in terms of  $\chi_6$ and $\chi_9$
the two choices yield equivalent but different looking
 solutions for the $4.5$PN
reactive terms in the equations of motion. 

Of the two choices, the second choice is more convenient for
calculations by hand since $ d{\bf J}/dt = 0$ to $O({\epsilon}^2)$,
but has no special advantage when the calculation is done on a computer.

\section{Redundant equations and related variant schemes}
\label{sec:altscheme}

It was noticed in IW that both at the 2.5PN
and at the 3.5PN order, the `balance procedure'
leads to two redundant constraint equations \cite{iw95}.
Here, at 4.5PN order, we once again obtain two redundant 
constraint equations. In this section, we examine critically 
the origin of these redundant equations. 

In implementing the `refined balance procedure' for the general
orbits, IW\cite{iw95} balance the `energy flux' and `angular momentum flux'
completely independently of each other. However, for circular 
orbits, these fluxes are not independent but related \cite{nishi} via: 
\[
\biggl( \frac{d{\cal E}}{dt} \biggr)_{\rm far-zone} = v^2\dot{\cal J}
\]
where $\dot{\cal J}$ is defined by the equation 
\[ 
\biggl( \frac{d\bf J}{dt} \biggr)_{\rm far-zone} = {\bf L}_N \dot{\cal J}
\]
The general balance should  reflect this limit and  we find
that for Newtonian RR a linear combination of the 
6 equations representing energy balance 
and another linear combination of the 6 equations representing
angular momentum balance are indeed identical and given by:
\begin{equation}
e_1 +e_2 -4 =0\,.
\end{equation}
Similarly at 3.5PN we have 
\begin{equation}
g_1 +g_2 +g_6 -\left(3 -\eta \right)\beta_2 
+{1\over 84}\left( 2927 -252\eta\right) =0\,,
\end{equation}
and finally at 4.5PN order the `degenerate' equation is
\begin{eqnarray}
k_1 +k_3 +k_5 +k_{10} + (3 -\eta)(\xi_1 +\xi_3) +
{1 \over 4}(90 +13\eta +6\eta^2)\beta_2 \nonumber \\
-{ 1\over 4536}(635771 +297117\eta -81000\eta^2) =&& 0
\,.
\end{eqnarray}
Thus we can trace the existence of
one of the redundant equations in the IW procedure
to the fact that for circular orbits the 
energy and angular momentum fluxes are not independent but
proportional to each other.

The mystery of the other redundant equation was not so easy
to resolve but after a careful examination of the system of equations
and `experiments' in modifying the system, we could finally  track it back 
to its source. The observation that this
redundant equation relates the coefficients of
the polynomial representing the ambiguity in $\bf\tilde J$ led us
to examine the functional form that IW proposed as the starting
ansatz for the calculation. A comparison of the functional forms for
the ambiguity in $\tilde E$ and $\bf\tilde J$  Eqs. (~(\ref{EJ*4.5})) 
reveals that indeed
IW assume a more general possibility for $\bf\tilde J$ 
than required. 
The ambiguity in angular momentum leads to terms more general  
than required  by the far-zone flux formula and time derivative
of the leading term using the reactive acceleration.
The absence of such terms in the far-zone flux then yields only the
trivial solution for these additional variables in $\bf\tilde J$, 
and the second redundant equation is just a homogeneous linear
combination of these trivial solutions.
         Thus the second redundant  equation in the IW scheme
is due to the fact that the IW scheme  --- extended here to 4.5PN order ---
is not a `minimal' one.  

To verify this `conjecture' we experimented with alternatives 
for the functional form that one assumes
as the starting expression for the ambiguity in $\tilde E$ and
$\bf\tilde J$ --- the  $2.5$PN, $3.5$PN and $4.5$PN order terms.
In the first instance, we replace the IW scheme --- labelled for clarity
of reference by IW21  ---  by the `minimal' variant in
Eq.~(\ref{EJ*4.5}) --- labelled by IW22. The notation IW21
indicates e.g., that $(m/r)^2 $ is pulled out in $\tilde E $
while only $(m/r)^1$ is pulled out in ${\bf\tilde J}$.
As explained above, 
the minimal choice for ${\bf \tilde J^*}$ is
obtained by pulling out the factor  
$(8/5) \eta {\bf \tilde L}_N  (m/r)^2  \dot r $ from
arbitrary terms in ${\bf \tilde J^*}$, rather than the factor
$(8/5) \eta {\bf \tilde L}_N  (m/r)  \dot r $ as in the
IW scheme for ${\bf \tilde J^*}$. This {\it reduces by one} the order of
the polynomial in $v^2$, $\dot{r}^2$, and $m/r$ that constitutes
the arbitrariness, and consequently 
implies  a reduction in the number of   variables that characterise
the ambiguity in $\bf\tilde J$ to  one for
 ${\bf \tilde J}_{2.5 }$, three in  ${\bf \tilde J}_{3.5}$
and six in ${\bf \tilde J}_{4.5}$. 
Thus in the IW22 scheme, at the 2.5PN level we have 6 variables
in the reactive acceleration, 3 variables determining the energy
ambiguity $\tilde E_{2.5}$ and 1 variable determining the ambiguity
in $\bf \tilde J_{2.5}$ i.e., 10 variables in all. The balance
equations lead to 9 equations 
--- 6 from energy and 3 from angular momentum ---
of which  8 are linearly
independent. In other words, {\it there is only one redundant equation}.
The linear system of 8 equations for 10 variables is then the same as
before  and leads to the IW21 solution in terms of 2 arbitrary parameters.
(The two extra variables in IW21 are identically zero.)
Similarly, at the 3.5PN level we have 12 variables
in the reactive acceleration, 6 variables determining the energy
ambiguity $\tilde E_{3.5}$ and 3 variables determining the ambiguity
in ${\bf \tilde J}_{3.5}$, i.e., 21 variables in all. The balance
equations lead to 16 equations ---  10 from energy and  6 from angular
momentum  --- of which  15 are linearly
independent, leaving {\it only one redundant equation}.
The linear system of 15 equations for 21 variables is then the same as before
and leads to the IW21 solution
in terms of 6 arbitrary parameters.
(The three extra variables in IW21 are identically zero.)
Finally, at the 4.5PN level, we have 20 variables
in the reactive acceleration, 10 variables determining the energy
ambiguity $\tilde E_{4.5}$ and 6 variables determining the ambiguity
in ${\bf \tilde J}_{4.5}$, i.e., 36 variables in all. The balance
equations lead to 25 --- 15 from energy and 10  from angular momentum ---
equations of which  24 are linearly
independent, again leaving {\it only one redundant equation}.
The linear system of 24 equations for 36 variables is the same as before
and leads to the solution obtained in the previous section 
in terms of 12 arbitrary parameters.
(The four extra variables in the IW21 scheme are identically zero.)
The IW22 (minimal) scheme thus confirms the conjecture that the occurence
of the second redundant equation is special to the IW scheme (IW21)
and is related to the choice they make for the functional form
of the $\bf\tilde J$ ambiguity by pulling out only one factor
of nonlinearity $m/r$ rather than its square --- the minimal choice.
To double check the above explanation, we performed another experiment
by examining a variant that would generate an increased  number of
redundant or degenerate equations. This  scheme denoted by IW11 differs from 
IW21  in that the ambiguity  in ${\bf \tilde E^*}$  is assumed to have
$(8/5) \eta  (m/r)  \dot r $ as the common factor, i.e.,
by pulling out only one order of nonlinearity $m/r$ rather than
its square as in IW21; the polynomial representing the ambiguity in
$\tilde E$ is consequently of {\it one order more} than in IW21. In this case,
at 2.5PN order one has $6+6+3=15$ variables and $10+6=16$ equations of which
{\it 3 are redundant}. The 13 equations for 15 variables thus yield
the required solution in terms of 2 arbitrary parameters and similarly
for higher orders. One may also explore the most general of choices
in which only $(8/5)\eta$ is pulled outside and the ambiguity is
the highest order polynomial consistent with the order of the
approximation. We studied one such scheme (IW00) in the Newtonian RR case. 
For convenience, the various experiments  are summarized
in Table \ref{tab:altscheme}.

To conclude: at 2.5PN, 3.5PN and 4.5PN orders all variants of
IW examined in this subsection 
with different forms of the ambiguities in $\tilde E$ and
$\bf\tilde J$ --- minimal (IW22) or IW11 --- lead to identical reactive
accelerations including their gauge arbitrariness.
 
At this juncture one may wonder about the issues of the `uniqueness'
and `ambiguities' of the schemes discussed earlier. In this regard, we
would like to make the following general remarks. For general orbits,
in addition to the balance of energy one must take into account the
balance of angular momentum. Thus, schemes involving only energy
balance are not relevant except in special cases like `circular
orbits' and `radial infall' (see section \ref{sec:qcho}). Can one have
schemes where one implements both energy and angular momentum balance
but does {\it not} take into account the possible ambiguities in
$\tilde E$ and $\bf\tilde J$? One can show that even at the 2.5PN
level this system of equations is inconsistent! Further, is the
ambiguity necessary {\it both } in $\tilde E$ and $\bf\tilde J$ ? If
one examines a scheme with both energy and angular momentum balance
taking account of the ambiguity {\it only} in $\tilde E$ one does
obtain a consistent solution upto 4.5PN order but with only half the
number of arbitrary parameters as in the IW scheme. The reduced
`gauge' freedom is not adequate to treat as special cases the
Burke-Thorne gauge at the 2.5PN level or the Blanchet choice at the
3.5PN level.  And finally, in a scheme with both energy and angular
momentum balance taking account of the ambiguity {\it only} in $\bf
\tilde J$ one obtains a consistent solution at 2.5PN order containing
{\it no} arbitrary parameters at all. No solution is possible at
higher orders.

On general considerations, the reactive acceleration should be a power
series in the individual masses $m_1$ and $m_2$ or equivalently, it
should be nonlinear in the total mass $m$ as assumed in earlier
sections. It is interesting to investigate whether the functional
forms of the far-zone fluxes and the balance procedure necessarily
lead to such `physical' solutions alone or whether they are consistent
with more general possibilities.  In the Appendix, for mathematical
completeness\cite{nonlin} we investigate this question in detail and
prove that the flux formulas and balance equations do not constrain
the reactive acceleration to their `physical' forms alone but allow
for a more general form for the reactive acceleration.

\section{Arbitrariness in reactive terms and gauge choice}
\label{sec:gauge}

It is well known that the formulas for the energy and angular momentum
fluxes in the far-zone are gauge invariant, i.e., independent of the
changes in the coordinate system that leave the spacetime
asymptotically flat. On the other hand, the expressions for the
reactive force are `gauge dependent' and consequently e.g., the
Chandrasekhar form is different from the Burke-Thorne or
Damour-Deruelle forms. In IW it was shown that the Burke-Thorne gauge
corresponds to the values $\beta_2=4 $ and $\alpha_3 =5$, while the
Damour-Deruelle choice corresponds to $ \beta_2 =-1 $ and $\alpha_3
=0$.  It was further shown that the reactive acceleration implied by
Blanchet's first principles determination of the 1PN radiation
reaction indeed corresponds to a particular choice of the arbitrary
parameters in the IW solution.  One of the satisfactory aspects of IW
was the demonstration that the part of the reactive acceleration not
determined by the balance requirement was precisely related to the
possible ambiguity in the choice of the gauge at that order. (The
flux is equal to the time variation of the conserved quantities only
upto total time derivatives; this ambiguity may be absorbed in a
`change' in the relative separation vector as discussed below.)
 
Following IW, we seek to establish the correspondence between the
arbitrary parameters contained in the radiation reaction terms and the
residual gauge freedom in the construction.  The residual gauge
freedom arises from the fact that the far-zone fluxes
Eqs. (~\ref{flux-schem}), (\ref{2PNfluxes}) are independent of changes in
the coordinate system that leave the spacetime asymptotically flat.
These coordinate changes will induce a change in $\bf x$ which is the
difference between the centers of mass of the two bodies ${\bf
  x}_1(t)$ and ${\bf x}_2(t)$ at coordinate time $t$. Following IW, we
choose the transformation to be of the form ${\bf x} \to {\bf x}'=
{\bf x} + \delta {\bf x}$, where $\delta {\bf x}$ can depend only on
the two vectors $\bf x$ and $\bf v$,
\begin{equation}
\delta {\bf x} = (f_{2.5} + f_{3.5} + f_{4.5}) \dot r {\bf x}  +
(g_{2.5}+g_{3.5}+g_{4.5}) r {\bf v} \,.
\label{deltax}
\end{equation}
In order that $\delta {\bf x} /{\bf x}$ be $O(\epsilon^{2.5})$,
$O(\epsilon^{3.5})$ and $O(\epsilon^{4.5})$, $f_{2.5}$ and $g_{2.5}$ 
must be $O(\epsilon^2)$, $f_{3.5}$ and $g_{3.5}$ must be
$O(\epsilon^3)$ and $f_{4.5}$ and $g_{4.5}$ must be $O(\epsilon^4)$.
As for the other variables, the $f$'s and $g$'s will also be polynomials
in the variables $m/r$, $v^2$ and $\dot r^2$. As
pointed out in \cite{iw95}, we do not independently take into account
changes in the coordinate time $t$ since the $\bf v$-dependent 
term in $\delta\bf x$ includes this contribution via $\bf x(t+\delta t)
\sim {\bf x}(t) +{\bf v}\delta t$.

In \cite{iw95} it was proved
 that to cancel the dependence on the two 2.5PN arbitrary
parameters and the six 3.5PN arbitrary parameters,
 $\delta {\bf x}$ should be chosen such
that
\begin{mathletters}
\label{1PNgaugef}
\begin{eqnarray}
f_{2.5} &=& {8\over15}\eta\left({m\over r}\right)^2\alpha_3 \,, \\
g_{2.5} &=& {8\over15}\eta\left({m\over r}\right)^2(2\alpha_3-3\beta_2)\,,\\
f_{3.5} &=& {8\over5}\eta\left({m\over r}\right)^2\biggl[ P_{21} v^2
+ P_{22}\left({m \over r}\right) + P_{23} \dot r^2 \biggr ]\,,\\
g_{3.5} &=& {8 \over 5} \eta\left({m \over r}\right)^2 \biggl[ Q_{21}v^2
+ Q_{22}\left({m \over r}\right) + Q_{23} \dot r^2 \biggr ]\,,
\end{eqnarray}
\end{mathletters}
where $P_{ab}$'s and $Q_{ab}$'s are given by
\begin{mathletters}
\label{3.5PNgauge} 
\begin{eqnarray}
P_{21} &=&{1 \over 3} \left[\xi_2+{2 \over 5}\xi_4-\rho_5-{1 \over 2} \alpha_3
(1-3\eta)\right]\,, \\
P_{22} &=&-{1 \over 6}\left[\xi_2+\xi_4-{3 \over 2} \xi_5-\rho_5 -
{3 \over 2}\beta_2 \eta +{1 \over 2} \alpha_3 (4+11\eta)\right]\,,\\
P_{23} &=&{1 \over 5} \xi_4  \,,\\
Q_{21} &=& \left[ \xi_1+{2 \over 3}\xi_2+{8 \over 15}\xi_4
+{1 \over2}(3\beta_2-2\alpha_3)(1-3\eta)\right]\,,\\
Q_{22} &=&-{1 \over 6} \left[6\xi_1+5\xi_2-3\xi_3 +5\xi_4-{3 \over
2}\xi_5+\rho_5-{63 \over 2}\beta_2 \eta -{1 \over
2}\alpha_3(4-55\eta)\right]\,,\\
Q_{23} &=& {1 \over 3} \left[{2 \over
5}\xi_4+\rho_5-\alpha_3(1-3\eta)\right]\,.
\end{eqnarray}
\end{mathletters}

We provisionally choose the 4.5PN part of $\delta {\bf x}$ to be of
the form
\begin{mathletters}
\label{2PNgauge-form}
\begin{eqnarray}
f_{4.5} &=& {8\over5}\eta \left({m \over r}\right)^2
              \left[P_{41}v^4+P_{42}v^2{m \over r}+P_{43}v^2 \dot r^2+
                P_{44}\left({m \over r}\right)^2+
                P_{45}\left({m \over r}\right) \dot r^2+
                P_{46}\dot r^4\right] \,, \\
g_{4.5} &=& {8\over5}\eta \left({m \over r}\right)^2
               \left[ Q_{41}v^4+Q_{42}v^2{m \over r}+Q_{43}v^2 \dot r^2+
                Q_{44}\left({m \over r}\right)^2+
                Q_{45}\left({m \over r}\right)\dot r^2+
                Q_{46}\dot r^4\right]\,.
\end{eqnarray}
\end{mathletters}
The change in the 2PN equations of motion Eqs.~(\ref{aPN}) produced by
this change of variable Eq.~(\ref{deltax}) can be determined using the
known form of $\delta \bf x$ upto 3.5PN order 
Eqs.~(\ref{1PNgaugef}), (\ref{3.5PNgauge}),
the provisional form chosen above for the 4.5PN
terms Eq.~(\ref{2PNgauge-form}) and the transformations given below:
\begin{eqnarray}
{\bf x} { \rightarrow} {{\bf x}'}&=& {\bf x} + \delta {\bf x}\,,\nonumber \\
{\bf v} { \rightarrow} {{\bf v}'}&=& {\bf v} +
\delta {\bf v}={d {\bf x} \over dt}+{d{\delta {\bf x}}\over dt}\,,\nonumber \\
r { \rightarrow}r'&=& r\left[ 1
 +{{\bf n}\cdot\delta {\bf {x}}\over r}\right]\,,\nonumber\\
{{\bf x}'\over r'^p}&=& {{\bf x}\over r^p} +{ \delta {\bf{x}}\over r^p}
- {p {\bf{n}}\over r^p}(\bf{n}\cdot\delta {\bf x})\,,\nonumber \\
v^2 {\rightarrow}v'^2 &=& v^2 +
\left[ 2{\bf v}\cdot{d {\delta {\bf x}}\over  dt}\right]\,,\nonumber\\
\dot r {\rightarrow} \dot r'&=& {1 \over r}\left[ r\dot r +
\delta {\bf x}\cdot{\bf v} + {\bf x}\cdot{d{\delta {\bf x}}\over dt}
-({\bf n}\cdot{\delta {\bf x}} )\dot r \right]\,.
\end{eqnarray}
The gauge change generates reactive terms 
and the requirement that this change should cancel the
dependence of the radiation-reaction terms on arbitrary parameters
dictates that 
\begin{mathletters}
\label{2PNgauge}
\begin{eqnarray}
P_{41} = && -{1\over24}\alpha_3(1-9\eta+21\eta^2)
          -{1\over30}(5\xi_2+2\xi_4-5\rho_5)(1-3\eta)\nonumber\\
         && +{1\over3}\psi_2 + {2\over15}\psi_4 + {8\over105}\psi_7
          -{1\over3}\chi_6 - {2\over15}\chi_8\\
P_{42} = && -{1\over6}\alpha_3(3+\eta^2) + {3\over8}\beta_2\eta
          -{1\over4}\xi_1\eta + {1\over12}\xi_2(3-23\eta)
          +{1\over60}\xi_4(19-77\eta) - {1\over8}\xi_5(1-3\eta) \nonumber\\
       && -{1\over12}\rho_5(3-22\eta) - {1\over3}\psi_2
          -{4\over15}\psi_4 + {1\over4}\psi_6 - {1\over5}\psi_7
          +{1\over12}\psi_8 + {1\over3}\chi_6 + {4\over15}\chi_8
          -{1\over4}\chi_9\\
P_{43} = && -{1\over10}\xi_4(1-3\eta) + {1\over5}\psi_4
          +{4\over35}\psi_7 - {1\over5}\chi_8\\
P_{44} = &&  {1\over30}\alpha_3(13+12\eta+16\eta^2)
          +{1\over5}\beta_2(1+12\eta-2\eta^2)
          +{1\over10}(\xi_1-2\xi_3)\eta\nonumber\\
         &&  +{1\over30}(\xi_2+\xi_4)(9+31\eta)
         -{1\over20}\xi_5(7+13\eta) 
         - {1\over30}\rho_5(9+28\eta)\nonumber\\
         && +{2\over15}\psi_2 + {2\over15}\psi_4 -{1\over10}\psi_6
          +{2\over15}\psi_7 - {1\over10}\psi_8 + {1\over5}\psi_9\nonumber\\
       && -{2\over15}\chi_6 - {2\over15}\chi_8 + {1\over10}\chi_9\\
P_{45} = &&  {1\over12}\alpha_3\eta^2 - {1\over4}\beta_2\eta(1-3\eta)
          +{1\over6}\xi_2\eta - {1\over15}\xi_4(1+7\eta)
          -{1\over3}\rho_5\eta\nonumber\\
         && - {1\over15}\psi_4
          -{2\over15}\psi_7 + {1\over6}\psi_8 + {1\over15}\chi_8\\
P_{46} = &&  {1\over7}\psi_7\\
Q_{41} = &&  {1\over8}(2\alpha_3-3\beta_2)(1-\eta-11\eta^2)
          -{1\over10}(15\xi_1+10\xi_2+8\xi_4)(1-3\eta)\nonumber\\
         &&  + \psi_1
          +{2\over3}\psi_2 + {8\over15}\psi_4 + {16\over35}\psi_7\\
Q_{42} = && -{1\over24}\alpha_3(108-331\eta+197\eta^2)
          +{1\over8}\beta_2(48-121\eta+63\eta^2)
          +{1\over2}\xi_1(9-28\eta) \nonumber\\
         &&+ {1\over12}\xi_2(49-142\eta)
           -{1\over8}(6\xi_3+3\xi_5-2\rho_5)(1-3\eta)
          +{1\over60}\xi_4(231-653\eta) \nonumber\\
         && - 2\psi_1 - {5\over3}\psi_2
          +{1\over2}\psi_3 - {47\over30}\psi_4 + {1\over4}\psi_6
          -{22\over15}\psi_7 + {1\over6}\psi_8 - {1\over6}\chi_6
          -{1\over10}\chi_8\\
Q_{43} = &&  {1\over6}\alpha_3(1-3\eta-3\eta^2)
          -{1\over6}(2\xi_2+2\xi_4+\rho_5)(1-3\eta)\nonumber\\
         && +{2\over15}\psi_4 + {16\over105}\psi_7
          +{1\over3}\chi_6 + {2\over15}\chi_8\\
Q_{44} = &&  {1\over30}\alpha_3(32+73\eta+254\eta^2)
          -{1\over30}\beta_2(51+157\eta+258\eta^2)
          -{1\over30}\xi_1(10-307\eta)\nonumber\\
         && -{1\over30}\xi_2(19-279\eta)
          -{1\over30}\xi_3(15+59\eta)
          -{1\over30}\xi_4(19-279\eta) 
          - {1\over60}\xi_5(9+91\eta)\nonumber\\
         && +{1\over30}\rho_5(9+28\eta) + {4\over3}\psi_1
           +{6\over5}\psi_2 - {1\over3}\psi_3 + {6\over5}\psi_4
          +{1\over3}\psi_5 - {7\over30}\psi_6 \nonumber\\
         && + {6\over5}\psi_7
          -{7\over30}\psi_8 + {2\over15}\psi_9 + {2\over15}\chi_6
          +{2\over15}\chi_8 - {1\over10}\chi_9\\
Q_{45} = && -{1\over24}\alpha_3(24-29\eta-91\eta^2)
          -{33\over8}\beta_2\eta^2 + {3\over4}\xi_1\eta
          +{1\over12}\xi_2(2+\eta) + {1\over15}\xi_4(6-13\eta)\nonumber\\
       && -{1\over4}\xi_5(1-3\eta) - {3\over4}\rho_5\eta
          -{1\over10}\psi_4 - {1\over5}\psi_7 + {1\over12}\psi_8
          -{1\over6}\chi_6 - {7\over30}\chi_8 + {1\over4}\chi_9\\
Q_{46} = && -{1\over5}\xi_4(1-3\eta) + {2\over35}\psi_7
          +{1\over5}\chi_8
\end{eqnarray}
\end{mathletters}
The above computation shows that as at the 3.5PN order the 
(12 parameter) arbitrariness
in the 4.5PN radiation reaction formulas reflects the residual freedom 
that is available to one in the choice of a 4.5PN accurate `gauge'. Every
particular 4.5PN accurate radiation reaction formula should correspond
to a particular choice of these 12 parameters.

\section{Particular cases: quasi-circular orbits and head-on infall}
\label{sec:qcho}

In this section we specialise our solutions valid for general orbits
to the particular case of quasi-circular orbits and radial infall and
verify that they indeed reproduce the simpler reactive solutions one
would obtain if one formulated the problem {\it ab initio} appropriate to
these two special cases.  We first consider the quasi-circular limit
that is of immediate relevance to sources for the ground based interferometric
gravitational wave detectors.  In this particular case, the reactive
acceleration may be deduced using {\it only} the energy balance.
Using the reactive acceleration we compute the 4.5PN contribution to
$\dot r$ and $\dot \omega$.  We also discuss the complementary case of
the radial infall of two compact objects of arbitrary mass ratio and
determine the 4.5PN contribution to the radial infall velocity for the
two special cases: radial infall from infinity and radial infall with
finite initial separation.

\subsection{Quasi-circular inspiral}

Using our general reactive solution we can 
compute the physically relevant quantities
$ \dot{r} $  and $ \dot{\omega} $ for quasi-circular inspiral,
where $ r $ and $
\omega $ are the orbital separation and the orbital angular frequency
in harmonic coordinates, respectively.  As would be expected,
these results are independent of the arbitrary parameters that are present
in the reactive solution.
We obtain the radiation reaction contribution to $\bf a $ upto 4.5PN
for quasi-circular inspiral by setting $\dot r =0 +O(\epsilon^{2.5}) $ 
and using 
\begin{equation}
v^2 = {m\over r}\left[1 - (3 -\eta)\frac{m}{r}
+\left( 6 +{41 \eta\over 4} +\eta^2\right)
\left(\frac{m}{r}\right)^2\right]
\label{eq:circv}
\end{equation}
in Eqs.~(\ref{provisional}), (\ref{ABexp}), (\ref{AB2PN}) and (\ref{sol4.5}).
We get
\begin{eqnarray}
{\bf a}_{RR} &=& -{32\eta m^3{\bf {v}}\over 5 r^4}\biggl[ 1 
-\left( {3431\over 336}-{5\over 4}\right)\frac{m}{r}\nonumber\\
&&\qquad
+\left( {794369\over 18144} +{26095\over 2016}\eta -{7\over 4}\eta^2
\right)\left(\frac{m}{r}\right)^2\biggr]\,.
\end{eqnarray}
It is worth noting that for quasi-circular inspiral
the energy flux determines the reactive
acceleration without any gauge ambiguity. 
All the  arbitrary terms in  energy are proportional
to $\dot r$ and hence play no role in this instance.
Inverting Eq.~(\ref{eq:circv}), we get
\begin{equation}
{m \over r} = v^2\left [ 1 +(3 -\eta)v^2 
            +{1 \over 4}( 48 -89\eta +4\eta^2)v^4 \right ]
\label{rveqn}
\end{equation}
Differentiating Eq. (\ref{rveqn}) w.r.t $t$ and noting that the {\bf a}
that appears is the total acceleration ( conservative + reactive)
we get, after some rearrangement
\begin{eqnarray}
\label{eq:rdot}
\dot{r} &=& -\frac{64}{5}\eta\left(\frac{m}{r}\right)^3
            \biggl[1-\left(\frac{1751}{336}
              +\frac{7\eta}{4}\right)\frac{m}{r} \nonumber\\
&&\qquad
              +\left(\frac{303455}{18144}+\frac{40981\eta}{2016} 
              +\frac{\eta^2}{2}\right)\left(\frac{m}{r}\right)^{2}
            \biggr]\,.
\end{eqnarray}

Using Eq.~(\ref{eq:rdot}) and the expression for angular velocity
($\omega\equiv v/r$)
\begin{equation}
\omega^{2} = {m\over r^3}\biggl[1-(3-\eta)\frac{m}{r}
 +( 6+\frac{41\eta}{4}+\eta^2)\left(\frac{m}{r}\right)^2 \biggr]\,,
\end{equation}
we may express $\dot\omega$ as 
\begin{eqnarray}
\label{eq:omegadot}
{\dot{\omega}\over \omega^{2}}&=& 
  \frac{96}{5}\eta(m\omega)^\frac{5}{3}
    \biggl[1-(m\omega)^\frac{2}{3}
      \left(\frac{743}{336} + \frac{11}{4}\eta\right)\nonumber\\
&&\qquad
  +\left(\frac{34103}{18144} +\frac{13661}{2016}\eta
   +\frac{59}{18}\eta^2\right)
   (m\omega)^\frac{4}{3}\biggr]\,.
\end{eqnarray}

The results Eqs.~(\ref{eq:rdot}) and (\ref{eq:omegadot}) 
are in agreement with \cite{BDIWW} as expected and required,
suggesting that the reactive terms obtained here
could be used to evolve orbits in the more general case also \cite{Orbit} .

\subsection{Head-on infall}

Recently Simone, Poisson and Will\cite{SPW95} have obtained to 2PN
accuracy the gravitational wave energy flux produced during head-on infall
and starting from these formulas one can deduce
{\it ab initio} the reactive acceleration in this limit adapting IW
to the radial infall case. As required, 
these results match 
exactly with expressions obtained by applying  radial infall limits  to
the general orbit solutions and we summarize the relevant formulas
in this limit in what follows.
Equations representing the head-on infall can be obtained 
from the general
orbit expressions by imposing the restrictions,
$ {\bf x}= z\hat{\bf n}$, ${\bf v}=\dot{z}\hat{\bf n}$,
$r=z$ and $v=\dot{r} = \dot{z} $.
For radial infall the conserved energy Eq.~(\ref{EJ}) 
to 2PN order then becomes
\begin{eqnarray}
E(z) &=& \mu\left\{ {\dot{z}^2\over 2} -\gamma 
+ {3\left( 1 -3\eta\right)\dot{z}^4
\over 8} +{ \left(3 +2\eta\right)\gamma\dot{z}^2\over 2}
+{\gamma^2\over2} \right.\nonumber \\
& &\left. + {5\left( 1 -7\eta +13\eta^2\right)\dot{z}^6\over 16}
+{3\left(7 -8\eta -16\eta^2\right)\gamma\dot{z}^4\over8}+\right.\nonumber\\
& &\left.{\left( 9 + 7\eta +8\eta^2\right)\gamma^2\dot{z}^2\over4}
-{\left( 2 +15\eta\right)\gamma^3\over4}\right\}\,,
\label{rf1}
\end{eqnarray}
where $ \gamma = m/z $. Unlike the quasi-circular inspiral, 
for head-on infall we can distinguish between  two different cases.
Following \cite{SPW95} we denote them by (A) and (B), respectively, 
and list the expressions relevant for our computations.
In  case (A), the radial infall proceeds from rest at infinite 
initial separation,  
$E(z)= E(\infty)=0$, and inverting Eq.~(\ref{rf1}) we get
\begin{equation}
\dot{z} = -\left\{ {2m\over z}\left[ 1 - 5{\gamma}
\left(1 -\frac{\eta}{2}\right)
+{\gamma^{2}}\left( 13 -\frac{81\eta}{4} +5\eta^{2}\right)
\right]\right\}^{1/2}\,.
\label{rf2}
\end{equation}
In case (B), the radial infall proceeds from rest at 
finite initial separation
$z_{0}$, which implies
\begin{equation}
E(z) = E(z_{0}) = -\mu\left\{ \gamma_{0} -{\gamma_{0}^2\over2}
+{\gamma_{0}^{3}\over2}\left( 1 +\frac{15\eta}{2}\right)\right\}\,.
\end{equation}
We obtain as in case (A), an expression for $\dot z$ given by
\begin{eqnarray}
\dot{z} &=& -\left\{ 2\left(\gamma -\gamma_{0}\right)\left[ 1
 -5{\gamma}\left(1 -\frac{\eta}{2}\right)
+{\gamma_{0}}\left(1 -\frac{9\eta}{2}\right)\right.\right.\nonumber\\
& &\left.\left.+{\gamma^{2}}\left( 13 -\frac{81\eta}{4} 
+5\eta^{2}\right)
-{\gamma\gamma_{0}}
\left( 5 -\frac{173\eta}{4} +13\eta^2\right)
\right.\right.\nonumber\\
& &\left.\left. +{\gamma_{0}^{2}}
 \left( 1 -\frac{5\eta}{4} + 8\eta^2\right)
\right]\right\}^{1\over2}\,,
\label{rf3}
\end{eqnarray}
where $\gamma_0= {m/z_0}$. 
We first compute the 4.5PN contribution to $\ddot z$ for case (B),the radial
infall from finite initial separation.
We use the radial infall restriction along with Eq.~(\ref{rf3}) in
Eqs.~(\ref{provisional}), (\ref{ABexp}), (\ref{AB2PN}) and (\ref{sol4.5})
to obtain 4.5PN terms in $\ddot z$ as 
\begin{eqnarray}
\ddot z &=& {8\eta\gamma^3 \over 5m }(2\gamma-2\gamma_0)^{1\over2}
  \biggl\{ {1\over 3}(-41 +21\zeta_1)\gamma+(8-4\zeta_1)\gamma_0\nonumber\\
    &&+\biggl[\biggl(\frac{1}{84}(18054-13231\eta)
    -{1\over 4}(438 -331 \eta)\zeta_1 + 18\zeta_2 
    +9\zeta_3 \biggr)\gamma^2 \nonumber\\
    &&+\biggl(-{1\over28}(5510-8849\eta)
    +{1\over4}(402-643\eta)\zeta_1 -26\zeta_2 -6\zeta_3\biggr)
    \gamma\gamma_0\nonumber\\
    &&+\biggl(36-126\eta-(18 -63\eta)\zeta_1+8\zeta_2\biggr)
    \gamma_0^2 \biggr]\nonumber\\
    &&+\biggl[\biggl(-\frac{1}{18144}
    (30549820 - 54233376\eta + 15776427\eta^2)\nonumber\\
    &&+{1\over32}(27156-49816\eta+15057\eta^2)\zeta_1\nonumber\\
    &&  
    -{1\over2}(766-527\eta)\zeta_2
    -{1\over4}(546-417\eta)\zeta_3
    +22\zeta_4+44\zeta_5 + 11\zeta_6 \biggr) \gamma^3\nonumber\\
    &&
    +\biggl( \frac{1}{3024}(6314916 - 20766190\eta + 8663249\eta^2)
    -{1 \over 16}(17052 -56198 \eta +23811\eta^2)\zeta_1\nonumber\\
    &&
    +(680 -759\eta)\zeta_2
    +{1 \over 4}(546 -855\eta)\zeta_3 -34\zeta_4 
    -104 \zeta_5 -8\zeta_6 \biggr)\gamma^2\gamma_0 \nonumber\\
    &&
    +\biggl( -{1\over2016}(1521308-7938232\eta+5800187\eta^2)
    +{1\over32}(12372 - 64104\eta +46641\eta^2)\zeta_1 \nonumber\\
    &&
    -{1\over2}(682-1315\eta)\zeta_2 
    -{1 \over 2}(54-189\eta)\zeta_3 +
    12\zeta_4 +76\zeta_5\biggr)\gamma\gamma_0^2 \nonumber\\
    &&
    +\biggl({1\over 8}(348-2016\eta+3339\eta^2)(2-\zeta_1)
    +( 44-162\eta)\zeta_2-16\zeta_5\biggr)\gamma_0^3 \biggr]
\biggr\}
\label{rf5}
\end{eqnarray}
To obtain the  2PN reactive
terms for case (A), the radial infall from infinity, we use in
Eqs.~(\ref{provisional}),~(\ref{ABexp}),~(\ref{AB2PN}) and (\ref{sol4.5})
the radial infall restriction and Eq.~(\ref{rf2}). The expression
thus obtained is the same as obtained by putting 
$\gamma_0 =0$ in Eq.~(\ref{rf5}).
The  $ \zeta $ 's in Eq.~(\ref{rf5}) are given by
\begin{eqnarray}
\zeta_1&=& \alpha_3 -\beta_2 \,,\nonumber\\
\zeta_2&=& \xi_1 +\xi_2+\xi_4 \,,\nonumber\\
\zeta_3&=& \xi_3 +\xi_5 \,,\nonumber\\
\zeta_4&=& \psi_3 +\psi_6+ \psi_8\,,\nonumber\\
\zeta_5&=& \psi_1 +\psi_2+ \psi_4 +\psi_7\,, \nonumber\\
\zeta_6&=&\psi_5 +\psi_9\,.
\end{eqnarray}
We have also computed the $2$PN
reactive terms for cases (A) and (B) {\it ab initio} using the IW
method adapted to radial infall.  
In this case, only 
energy balance is needed as ${\bf J} =0$ for head-on infall.
The result thus obtained is  in agreement with Eq.~(\ref{rf5}).
Eq.~(\ref{rf5})  may be integrated straightforwardly  
to obtain the $4.5$PN contribution to
$\dot z^2 $ in case (B) and it yields
\begin{eqnarray}
\dot z^2 &=& {16(2\gamma -2\gamma_0)^{3 \over 2}\eta \over 5 }
\biggl\{ {1 \over 21}(41 -21\zeta_1)\gamma^2
 -{4 \over 105}\gamma\gamma_0 -{8\over 315}\gamma_0^2\nonumber\\
 &&+\biggl[\biggl({1 \over 756}(-18054 
+13231\eta)
+{ 1 \over 36}( 438 -331\eta)\zeta_1
-2\zeta_2 -\zeta_3\biggr)\gamma^3\nonumber\\
&& +\biggl( {1 \over 252}
( 1926 -7597\eta)
+{1 \over 168}(660 -2534\eta)\zeta_1
+2 \zeta_2\biggr)\gamma_0\gamma^2\nonumber\\
&&\biggl( {1 \over 315}( -342 +341\eta)
+{ 1\over 525}( 240 -280\eta)\zeta_1\biggr)\gamma_0^2\gamma\nonumber\\
&&+ \biggl({1 \over 945}( -684 +682\eta)
+{ 1\over
  4725}(1440-1680\eta)\zeta_1\biggr)\gamma_0^3\biggr]\nonumber\\
&&
+\biggl[\biggl( {1091065 \over 7128} -{564931\eta \over 2079}
+{ 5258809\eta^2 \over 66528} 
-{1 \over 352}(27156 - 49816\eta +15057\eta^2)\zeta_1 
\nonumber\\
&& 
+{1 \over 22}(766-527\eta)\zeta_2
+{1 \over 44}( 546 -417\eta)\zeta_3
-2 \zeta_4 -4\zeta_5 -\zeta_6\biggr)\gamma^4\nonumber\\
&&
+\biggl( -{ 21548237\over 224532} + {26019487 \eta \over 49896}
-{ 2750389 \eta^2\over 11088} + {1 \over 528}( 26316 - 139638\eta 
+67231\eta^2 )\zeta_1\nonumber\\
&&
-{1 \over 99}( 4416 -6241\eta)\zeta_2
-{1 \over 132}( 546 -2023\eta)\zeta_3
+2\zeta_4 +8\zeta_5 \biggr)\gamma_0\gamma^3\nonumber\\
&&+\biggl( {3823453 \over 149688} -{1681430\eta \over 14553}
+{ 4399627 \eta^2\over 22176} -{1 \over 2464}( 30828 -146592\eta
+244127\eta^2)\zeta_1 \nonumber\\
&&
+{ 1\over 5082}( 53262 -202741\eta)\zeta_2
+{ 1\over 77}( 24 -28\eta)\zeta_3 
-4\zeta_5\biggr)\gamma_0^2\gamma^2 \nonumber\\
&&+\biggl( {567739 \over 187110} +{608992\eta\over 72765}
-{ 228227 \eta^2\over 27720} -{1 \over 385}( 504 +
1080\eta -1622\eta^2)\zeta_1\nonumber\\
&& -{ 1\over 2541}(1056 -1232\eta^2)\zeta_2
+{1 \over 385}( 96 -112\eta)\zeta_3\biggr)\gamma_0^3\gamma
+\biggl( {567739\over 280665} +{1217984\eta \over 218295}\nonumber\\
&&
-{228227 \eta^2\over 41580} 
-{ 1\over 1155}(1008+2160\eta-3244\eta^2)\zeta_1
-{ 1 \over 22869}(6336 -7392\eta)\zeta_2\nonumber\\
&&
+{ 1\over 63525}(10560 -12320\eta)\zeta_3\biggr)\gamma_0^4
\biggr]\biggr\} \,.
\label{zdotb}
\end{eqnarray}
We obtain the 4.5PN contribution to $ \dot z^2 $ for case (A) 
by putting $ \gamma_0=0$ in Eq.~(\ref{zdotb}).
Unlike in the case of quasi-circular inspiral 
the expressions in the head-on or radial infall cases are 
dependent on the choice of arbitrary variables or the choice
of `gauge'.

\section{Concluding remarks}
\label{sec:conclusions}

Starting from the 2PN accurate energy and angular momentum fluxes for
structureless non-spinning compact binaries of arbitrary mass ratio
moving on quasi-elliptical orbits we deduce the 4.5PN reactive terms
in the equation of motion by an application of the IW method. The
4.5PN reactive terms are determined in terms of twelve arbitrary
parameters which are associated with the possible residual choice
of `gauge' at this order. These general results could prove useful
to studies of the evolution of the orbits.
The limiting and complementary cases of circular orbits and head-on infall
have also been examined. 

We have systematically and critically
explored different facets of the IW  choice like the
functional form of the reactive acceleration and provided a better
understanding of the origin of redundant equations  by studying 
variants obtained by modifying the functional forms of 
the ambiguities in $\tilde E^*$ and 
$\bf\tilde J^*$.
The main conclusions we arrive at by this analysis are
\begin{itemize}
\item In terms of the number of arbitrary parameters and the
  corresponding gauge transformations, the IW scheme exhibits
  remarkable stability for a variety of choices for the form of the
  ambiguity in energy and angular momentum. The different choices
  merely produce different numbers of degenerate equations. This
  indicates the essential validity and soundness of the scheme.
These solutions are general enough
to treat as special cases any
particular solutions obtained from first principles in the future.
\item Relaxing the requirement of nonlinearity in $m$ 
or more precisely the power series behaviour in $m _1 $ and $ m_2$ 
  permits
  mathematically more general solutions for the reactive 
    accelerations  involving more arbitrary
  parameters. 
 Solutions more general than the ones discussed
  in the appendix, e.g., a solution involving 6 parameters at the
  Newtonian level, cannot be gauged away either by gauge
  transformations of the form discussed by IW or by more general gauge
  transformations that differ in their powers of nonlinearity
  ($m/r$ dependence).
   However, none of these solutions are of `physical' interest
  to describe the gravitational radiation reaction of two-body 
  systems. 
\end{itemize}
\section*{Acknowledgments} 
We are particularly grateful to  Thibault Damour  for  discussions 
clarifying the nonlinear structure of the reactive acceleration ansatz. 
We thank Luc Blanchet, Clifford Will and Rajaram Nityananda 
for their valuable comments. 
One of us (SI) would like to thank the
Raman Research Institute for hospitality during the initial stages
of this collaboration.

\appendix
\section{The general solution to the balance method}

\subsection{The 2.5PN reactive solution}

It should be noted that all the discussion in section \ref{sec:altscheme}
follows only after one has {\it assumed} a functional form for the reactive
acceleration --- in particular, the intuitive requirement that 
it be nonlinear, i.e., contain an overall factor of $m/r$.
It is pertinent to ask whether more general possibilities
obtain, consistent with the far-zone fluxes, if one relaxes this
requirement. We have explored this question in  detail
at the 2.5PN level and we summarize the results in what follows.
In this instance the reactive acceleration is assumed to be:
\begin{eqnarray}
\label{provnew}
{\bf a}&=&-{8 \over 5} \eta\left(\frac{m}{r^2}\right)
\bigl[-({\cal A}_{2.5})\dot r
{\bf n} + ({\cal B}_{2.5}){\bf v}\bigr] \,,\nonumber\\
{\cal A}_{2.5}=&&a'_1v^4+a'_2v^2{m \over r}+a'_3v^2 \dot r^2+
a'_4\left({m \over r}\right)^2
+a'_5 \left({m \over r}\right)\dot r^2+a'_6 \dot r^4 \,, \nonumber\\
{\cal B}_{2.5} = &&b'_1v^4+b'_2v^2{m \over r}+b'_3v^2 \dot r^2+
b'_4 \left({m \over r}\right)^2
+b'_5 \left({m \over r}\right)\dot r^2+b'_6 \dot r ^4  \,,
\end{eqnarray}
i.e., it is determined by 12 reactive coefficients instead
of the earlier 6. 
Recall that the nomenclature IW22, IW21 and IW11 refers to
the functional forms chosen for the ambiguity in
energy and angular momentum and we introduce  similar notation
EJ22, EJ21 and EJ11, respectively,
in this appendix, 
where the acceleration has a more general form as given by Eq.~(\ref{provnew}).
With this form of the reactive acceleration, however, one 
gets  e.g., in the EJ21 scheme at $2.5$PN 
\begin{mathletters}
\label{EJ*5/2}
\begin{eqnarray}
\tilde E^* &&\equiv \tilde E_N + \tilde E_{2.5} \nonumber \\
&&= \tilde E_N -{8 \over 5} \eta \left({m \over r}\right) ^2
\dot r  ( \alpha_1 v^2 + \alpha_2 {m \over r} +\alpha_3 \dot r^2
) \,, \label{E*5/2} \\
{\bf \tilde J^*} &&\equiv {\bf \tilde L}_N + {\bf \tilde J}_{2.5}
\nonumber \\
&&={\bf \tilde L}_N + {8 \over 5} \eta {\bf \tilde L}_N  {m \over r}  \dot r
( \beta_1 v^2 + \beta_2 {m \over r} + \beta_3 \dot r^2 ) \,,
\end{eqnarray}
\end{mathletters}
The derivatives of $\tilde E^*$ and ${\bf \tilde J^*}$ with the new
form of the reactive acceleration are given by
\begin{mathletters}
\label{EJ*dot5/2}
\begin{eqnarray}
{{d \tilde E^*} \over {dt}}&=& -{8 \over 5} \eta {{m} \over r^2}
\biggl[ (b'_1)v^6 +(b'_2 +\alpha_1){m \over r}v^4 + ( -a'_1 +b'_3)\dot r^2 v^4
+( b'_4 -\alpha_1 +\alpha_2)\left({m \over r}\right)^2 v^2\nonumber\\
&&
  +( -a'_3 +b'_6)\dot r^4 v^2
+( -a'_2 +b'_5 -3 \alpha_1 +3\alpha_3)\left({m \over r}\right)\dot r^2 v^2
-\alpha_2\left({m \over r}\right)^3 \nonumber\\
&&
- (a'_4+2\alpha_1 +
4 \alpha_2 +3\alpha_3)\left({m \over r}\right)^2\dot r^2
-( a'_5 +5 \alpha_3)\left({m\over r}\right)\dot r^4 - a'_6\dot r^6
\biggr]\,, \\
{{d {\bf \tilde J^*}} \over {dt}}&=&-{8 \over 5} \eta {\bf \tilde L}_N
\left({ m\over r^2}\right)
\left[ (b'_1 -\beta_1)v^4 + (b'_2 +\beta_1 -\beta_2)
\left({m \over r}\right)v^2
+ ( b'_3 +2\beta_1 -3\beta_3)\dot r^2 v^2 \right.\nonumber\\
& &\left.
+( b'_4 +\beta_2)\left({m \over r}\right)^2 +( b'_5+2\beta_1
+3\beta_2 +3\beta_3)\left({m\over r}\right)\dot r^2
+( b'_6 +4 \beta_3)\dot r^4\right]\,.
\end{eqnarray}
\end{mathletters}
Using Eqs.~(\ref{EJ*5/2}) and (\ref{EJ*dot5/2}) one can understand the 
counts of the various variables summarised in Table \ref{tab:altascheme}.

One can explain the new counts for the arbitrary parameters by
comparing e.g., the EJ21 scheme with a general form for the reactive
acceleration as in this section with the IW21 scheme with the
restricted form for reactive acceleration as in section
\ref{sec:altscheme}.  One has 6 extra variables and 4 extra equations.
However one gains an extra equation because one of the degeneracies is
lifted. The resulting 5 equations for 6 variables lead to an extra
arbitrary parameter resulting in a 3 parameter solution in this
instance. All the other entries in Table \ref{tab:altascheme} can be
similarly understood by comparison of Tables \ref{tab:altscheme} and
\ref{tab:altascheme}.

The reactive solution resulting from the EJ22 scheme in this
instance is exactly the same as the IW21 reactive solution discussed
earlier.  From the  EJ21 scheme one obtains a  solution with three
arbitrary parameters given by
\begin{mathletters}
\label{nsnsol}
\begin{eqnarray}
&&
a'_1=3\beta_3 \,, \quad
a'_2=3(1 +\alpha_3 - \beta_3) \,, \quad
a'_3= -4 \beta_3 \,, \nonumber \\
&&
a'_4 = 23/3 -3\alpha_3 +2\beta_2 \,,
a'_5= -5\alpha_3 \,, \quad
a'_6=0\,, \\ 
&&
b'_1=0 \,, \quad
b'_2= 2 +\beta_2 \,, \quad
b'_3= 3\beta_3  \,, \quad
b'_4 = 2 -\beta_2 \,, \nonumber\\
&&
b'_5= -3\left( 1 +\beta_2 +\beta_3\right) \,, \quad
b'_6=-4 \beta_3\,.
\end{eqnarray}
\end{mathletters}
This construction can be generalised to 3.5PN and 4.5PN orders
in  which cases the number of arbitrary parameters are 8 and 15, respectively.
The EJ11 and EJ00  schemes on the other hand lead to
a solution with six arbitrary parameters at the 2.5PN level. 
However, not all these solutions are similar
in regard to the possibility of gauging away all the arbitrary
parameters they contain.

\subsection{The 2.5PN gauge arbitrariness}

We have also investigated the question  whether all the
extra arbitrary parameters appearing in
schemes with the general form of  reactive acceleration 
(See Table \ref{tab:altascheme}.) can be gauged away?
We find that at 2.5PN order, though this is possible with the 3 parameters
of the EJ21 scheme, it is not true for the 6 arbitrary
parameters  in the EJ11 and EJ00 schemes.
For this reason  the EJ11 and EJ00  schemes  are  not satisfactory and
we discuss them no further. We present here for the EJ21 scheme
details of the gauge calculation at 2.5PN order.
We choose $ \delta {\bf x}$ to be
\begin{equation}
\delta {\bf x} ={ 8\eta\over 5}({m \over r})\left(f'_{2.5}\dot r {\bf x}
+ g'_{2.5} r {\bf v}\right) \,,
\label{deltaxi}
\end{equation}
where  $ f'_{2.5} $ and $ g'_{2.5} $ are given by
\begin{eqnarray}
\label{nsNGF}
f'_{2.5}&=&  P'_{01}({m \over r}) +P'_{02} v^2
+P'_{03} \dot r^2  \,,\nonumber\\
g'_{2.5}&=&  Q'_{01}({m \over r}) +Q'_{02} v^2
+Q'_{03}\dot r^2 \,.
\end{eqnarray}
For the reactive acceleration given by Eqs.~(\ref{provnew})
and (\ref{nsnsol}) we obtain
\begin{mathletters}
\label{nsNGS}
\begin{eqnarray}
%Gopu's expressions
%P'_{01}&=&-{1 \over 3}\left(\alpha_3 -\beta_3\right) \,,\\
%P'_{02} &=& -{1 \over 2}\beta_2\,,\\
%P'_{03} &=&  0 \,,\\
%Q'_{01}&=& {1 \over 3}\left( 3\beta_2 -\beta_3 -2\alpha_3\right) \,,\\ 
%Q'_{02} &=& 0 \,,\\
%Q'_{03} &=& {1 \over 2}\beta_3\,. 
%P'_{01}&=&-{1 \over 3}\left(\alpha_3 -\beta_3\right) \,,\\
P'_{01}&=&{1 \over 3}(\alpha_3 -\beta_3) \,,\\
P'_{02} &=& {1 \over 2}\beta_3\,,\\
P'_{03} &=&  0 \,,\\
Q'_{01}&=& {1 \over 3}(2\alpha_3-3\beta_2+\beta_3) \,,\\ 
Q'_{02} &=& 0 \,,\\
Q'_{03} &=& -{1 \over 2}\beta_3\,. 
\end{eqnarray}
\end{mathletters}

The EJ21 scheme leads to a more general solution to the balance
equations, and as in IW all the arbitrary parameters that appear in
its solution can be associated with a residual choice of gauge. It has
been explored in detail upto 4.5PN and the results are summarised
below.  We list the new general reactive solutions and the
corresponding gauge transformations for the arbitrary parameters they
contain.  For brevity, the solutions are presented in the form : `New
solution' = `Old solution' + `Difference'.

\subsection{ The 3.5PN and 4.5PN reactive solutions}

The reactive acceleration is assumed to have the following general
form
\begin{eqnarray}
{\bf a}&=&-{8 \over 5} \eta \frac{m}{r^2}\bigl[-({\cal A}_{2.5}
+ {\cal A}_{3.5} +{\cal A}_{4.5})\dot r
{\bf n} + ({\cal B}_{2.5}+{\cal B}_{3.5}
+{\cal B}_{2.5} ){\bf v}\bigr],
\end{eqnarray}
with  ${\cal A}_{2.5}$ and ${\cal B}_{2.5} $ given in 
Eqs.~(\ref{provnew}) and (\ref{nsnsol})
and
$ {\cal A}_{3.5},~
{\cal B}_{3.5},~{\cal A}_{4.5}$ and ${\cal B}_{4.5}$ given by
\begin{mathletters}
\begin{eqnarray}
{\cal A}_{3.5}&=& f'_1v^6+f'_2v^4{m \over r}+f'_3v^4 \dot r^2+
f'_4 v^2\dot r^2 {m \over r}+
f'_5 v^2\dot r^4 \nonumber\\
&& +f'_6 v^2\left({m \over r}\right)^2 +f'_7{m \over r}\dot r^4
+ f'_8\left({m \over r}\right)^2\dot r^2 
+f'_9\left({m \over r}\right)^3 +f'_{10}\dot r^6 \,,\\
{\cal B}_{3.5} &=& g'_1v^6+g'_2 v^4{m \over r}+g'_3v^4 \dot r^2+
g'_4 v^2\dot r^2 {m \over r}+g'_5 v^2\dot r^4 \nonumber\\
&& +g'_6v^2\left({m \over r}\right)^2
+g'_7{m \over r}\dot r^4 
+ g'_8\left({m\over r}\right)^2 \dot r^2 
+ g'_9 \left({m \over r}\right)^3 + g'_{10}\dot r^6 \,, \\
{ \cal A}_{4.5} &=& h'_1v^8 + h'_2v^6\dot r^2 + h'_3v^6{m\over r} +
  h'_4v^4\dot r^4 + h'_5v^4\left({m\over r}\right)^2 \nonumber\\
&&
  + h'_6v^4\dot r^2{m\over r} +
  h'_7 v^2\dot r^6 + h'_8 v^2\dot r^4{m\over r}
  + h'_9 v^2\dot r^2\left({m\over r}\right)^2 +h'_{10} v^2\left({m\over
    r}\right)^3\nonumber\\
&& 
+h'_{11}\left({m \over r}\right)^4
+ h'_{12}\dot r^2 \left({m \over r}\right)^3 
+h'_{13}\dot r^4\left({m \over r}\right)^2
+h'_{14}\dot r^6{m \over r} +h'_{15}\dot r^8 \,, \\
{ \cal B}_{4.5} &=& k'_1v^8 + k'_2v^6\dot r^2 + k'_3v^6{m\over r} +
  k'_4v^4\dot r^4 + k'_5v^4\left({m\over r}\right)^2 \nonumber\\
&&
  + k'_6v^4\dot r^2{m\over r} +
  k'_7 v^2\dot r^6 + k'_8 v^2\dot r^4{m\over r}
  + k'_9 v^2\dot r^2\left({m\over r}\right)^2 
 + k'_{10}v^2 \left({m\over r}\right)^3 \nonumber\\
&&
+k'_{11}\left({m \over r}\right)^4
+ k'_{12}\dot r^2 \left({m \over r}\right)^3 
+k'_{13}\dot r^4\left({m \over r}\right)^2
+k'_{14}\dot r^6{m \over r} +k'_{15}\dot r^8\,.
\end{eqnarray}
\end{mathletters}

With this form of the acceleration we have at $3.5$PN
\begin{mathletters}
\label{EJ*dot3.5A}
\begin{eqnarray}
{{d \tilde E^*} \over {dt}}&=& -{8 \over 15} \eta {{m} \over r^2}
\biggl [
({m \over r})^2\biggl ( 12v^2 -11\dot r^2 \biggr )
+\sum_{i=1}^{15}{\cal R'}^{[3.5]}_i  {\cal Y}^{[4]}_i
\biggr ] \,,\\
{{d {\bf \tilde J^*}} \over {dt}}&=& -{8 \over 5} \eta
{\bf \tilde L}_N
{{m} \over r^2} \biggl [
{m \over r}\biggl ( 2v^2 +2{m \over r} -3 \dot r^2  \biggr )+
\sum_{i=1}^{10}{\cal S'}^{[3.5]}_i  {\cal Y}^{[3]}_i
 \biggr ]\,,
\end{eqnarray}
\end{mathletters}
where ${\cal Y}^{[4]}_i $ is given by Eqs.~(\ref{cy4}),
\begin{eqnarray}
{\cal Y}^{[3]}_i(i=1\ldots10)&=&
   \biggl [ v^6, v^4{m \over r}, v^4\dot r^2, 
   v^2\left({m\over r}\right)^2, 
 v^2{m \over r}\dot r^2, v^2\dot r^4,\left({ m\over r}\right)^3,
\left({m \over r}\right)^2\dot r^2, {m \over r}\dot r^4, \dot r^6 \biggr ]
\end{eqnarray}
and ${\cal R'}^{[3.5]}_i$, ${\cal S'}^{[3.5]}_i$
consist of corresponding linear combinations of the parameters involved.
Repeating the procedure explained in the text, the $3.5$PN reactive solution
obtained is:
\begin{mathletters}
\begin{eqnarray}
f'_1&=& -{3 \over 2}(1 -3\eta)\beta_3 -3\rho_2\,,\\
f'_2 &=& f_1 -{ 1\over 2}(21 +39 \eta)\beta_3 +3 \rho_2 \,,\\
f'_3 &=& 2(1 -3\eta)\beta_3 +4\rho_2 -5\rho_4 \,,\\
f'_4&=& f_3 +{1 \over 2 }( 56 +15\eta)\beta_3 +2\rho_2 +5 \rho_4\,,\\
f'_5&=& 6\rho_4\,,\\
f'_6 &=& f_2 + (21 +12\eta)\beta_3\,,\\
f'_7 &=& f_5 -4\eta\beta_3\,,\\
f'_8 &=& f_4 -3\eta\beta_3\,,\\
f'_9 &=& f_6\,,\\
f'_{10}&=&0\,,\\
g'_1 &=& 0\,,\\
g'_2 &=& g_1\,,\\
g'_3  &=& -{ 3\over 2}( 1 -3\eta)\beta_2 -3\rho_2\,,\\
g'_4 &=& g_3 -{ 1\over 2}(21 +33\eta)\beta_3 + 3\rho_2\,,\\
g'_5 &=& 2(1 -3\eta)\beta_3 +4\rho_2 -5\rho_4\,,\\
g'_6 &=& g_2\,,\\
g'_7&=& g_5 +{ 1 \over 2}( 56 +\eta)\beta_3 +2\rho_2 +5\rho_4\,,\\
g'_8 &=& g_4 +(21 +9\eta)\beta_3\,,\\
g'_9 &=& g_6\,,\\
g'_{10} &=& 6\rho_4\,,
\end{eqnarray}
\end{mathletters}
where $f_i$,~$g_i $ are given by Eqs.~(\ref{cdsolution}).
The solution corresponding to
Eqs.~(\ref{3.5amb}) remains identical.

Similarly at 4.5PN we have
\begin{mathletters}
\label{EJ*dot4.5A}
\begin{eqnarray}
{{d \tilde E^*} \over {dt}}&=& -{8 \over 15} \eta {m \over r^2}
\biggl [
\left({m \over r}\right)^2\biggl ( 12v^2 -11 \dot r^2 \biggr )
+ \left({m \over r}\right)^2\left\{\frac{1}{28}\left[(785 -
852\eta)v^4\right.\right.\nonumber\\
& &\left.\left.+2(-1487 + 1392\eta)v^2\dot{r}^2
 + 160(-17 +\eta) {m \over r}v^2
+ 3(687 - 620\eta)\dot{r}^4 \right.\right.\nonumber\\
& & \left.\left.+ 8 (367 - 15\eta){m \over r}\dot r^2
 +16(1 - 4\eta)\left({m \over r}\right)^2\right] \right\}
+\sum_{i=1}^{15}{\cal R}'^{[4.5]}_i  {\cal Y}^{[4]}_i
\biggr ]\,, \\
{{d {\bf \tilde J^*}} \over {dt}}&=& -{8 \over 5} \eta 
{\bf \tilde L}_N
{{m} \over r^2} \biggl [
{m \over r}\biggl ( 2v^2 +2{m \over r} -3 \dot r^2  \biggr )+
{m \over r}\left\{ {1 \over 84}
\left[ (307 - 548\eta)v^4
+ 6\left(-74 + 277\eta \right)v^2\dot{r}^2 \right.\right.\nonumber\\
&& - 4(58 + 95\eta){m \over r}v^2 
+ 3(95 - 360\eta)\dot{r}^4 + 2(372 + 197\eta)
{m \over r}\dot{r}^2 \nonumber \\
&& \left.\left.
+ 2(-745 + 2\eta)\left({m \over r}\right)^2 \right]\right\}
+\sum_{i=1}^{15}{\cal S}'^{[4.5]}_i  {\cal Y}^{[4]}_i
 \biggr ]\,,
\end{eqnarray}
\end{mathletters}
where 
\begin{eqnarray}
{\cal Y}^{[5]}_i (i=1\ldots21)&=&
   \biggl [ v^{10}, v^8{m \over r}, v^8\dot r^2, 
     v^6\left({m \over r}\right)^2, v^6{m \over r}\dot r^2, 
     v^6\dot r^4,
     v^4\left({m \over r}\right)^3, 
     v^4\left({m \over r}\right)^2\dot r^2,\nonumber\\
&&\quad
   v^4{m \over r}\dot r^4, v^4\dot r^6, 
     v^2\left({m \over r}\right)^4, 
     v^2\left({m \over r}\right)^3\dot r^2,
     v^2\left({m \over r}\right)^2\dot r^4, \nonumber\\
&&\quad
   v^2{m \over r}\dot r^6, 
    v^2\dot r^8, \left({m \over r}\right)^5, 
    \left({m \over r}\right)^4\dot r^2, 
    \left({m \over r}\right)^3\dot r^4,
    \left({m \over r}\right)^2\dot r^6, {m \over r}\dot r^8,
    \dot r^{10} \biggr ] 
\end{eqnarray}
and  $ {\cal Y}^{[4]}_i$ is given by Eq.~(\ref{cy4}).\\
Here ${\cal R'}^{[4.5]}_i$,${\cal S'}^{[4.5]}_i$
consist of  linear combinations of the parameters involved.
The $4.5$PN reactive solution reads as:
\begin{mathletters}
\label{4.5PNRSA}
\begin{eqnarray}
h'_1 &=& -{ 1\over 8}(3 -27\eta +63\eta^2)\beta_3 +{ 3\over 2}(1 -3\eta)\rho_2
-3\chi_2\,,\\
h'_2 &=& { 1\over 2}(1 -9\eta +21\eta^2)\beta_3 -2(1 -3\eta)\rho_2 +{ 5\over 2}
(1 -3\eta)\rho_4 +4\chi_2 -5\chi_4\,,\\
h'_3 &=& h_1 +{1 \over 8}(3 -207\eta +75\eta^2)\beta_3 
+{1 \over 2}(21 +39\eta)\rho_2 +3\chi_2\,,\\
h'_4 &=& -3(1 -3\eta)\rho_4 +6\chi_4 -7\chi_7\,,\\
h'_5&=& h_3 +(18 +96\eta +18\eta^2)\beta_3 -(21 +12\eta)\rho_2\,,\\
h'_6&=& h_2 -{1 \over 4}(24 -397\eta +95\eta^2)\beta_3 
-{1 \over 2}(70 -11\eta)\rho_2 
+{ 1\over 2}(35 
+65\eta)\rho_4 +4\chi_2 +5\chi_4\,,\\
h'_7 &=& 8\chi_7\,,\\
h'_8 &=& h_4 +{1 \over 8}( -353 +195\eta)\eta\beta_3 +\eta\rho_2 
-{1 \over 2}(84 +25\eta)\rho_4 +2\chi_4 +7\chi_7\,,\\
h'_9 &=& h_6 -{ 1\over 4}(260 +119\eta +30\eta^2)\beta_3 
-(14 +5\eta)\rho_2 -(35 +20\eta)\rho_4 \,,\\
h'_{10} &=& h_5 -{1 \over 4}(306 +489\eta +48\eta^2)\beta_3\,,\\
h'_{11} &=& h_{10}\,,\\
h'_{12} &=& h_9 -{1 \over 4}(12 +87\eta -24\eta^2)\beta_3\,,\\
h'_{13}&=& h_8 -{1 \over 2}(8 +49\eta +34\eta^2)\beta_3 
+2\eta\rho_2 +5\eta\rho_4\,,\\
h'_{14} &=& h_7 +6\eta[ (1 -3\eta)\beta_3 +\rho_4]\,,\\
h'_{15}&=& 0\,,\\
k'_1 &=&0\,,\\
k'_{2}&=& -{ 1\over 8}(3 -27\eta + 63\eta^2 )\beta_3 
+{3 \over 2}(1 -3\eta)\rho_2 -3\chi_2\,,\\
k'_3&=&k_1\,,\\
k'_4 &=&{ 1\over 2}(1 -9\eta +21\eta^2)\beta_3 -2(1 -3\eta)\rho_2 +{ 5\over 2}
(1 -3\eta)\rho_4 +4\chi_2 -5\chi_4\,,\\
k'_5 &=& k_3\,,\\
k'_6 &=& k_2 +{ 3\over 8}( 1 -81\eta +13\eta^2)\beta_3 
+ {1 \over 2}( 21 +33\eta)\rho_2 + 3\chi_2\,,\\
k'_7 &=& -3(1 -3\eta)\rho_4 +6\chi_4 -7\chi_7\,,\\
k'_8 &=& k_4 -{ 1\over 4}(24 -421\eta -\eta^2)\beta_3
-{1 \over 2}( 70 -25\eta)\rho_2 +{ 1\over 2}( 35 +55\eta)\rho_4 
+4 \chi_2 +5\chi_4 \,,\\
k'_9 &=& k_6 +{1 \over 4}(84 +525\eta +54\eta^2)\beta_3 
-(21 +9\eta)\rho_2\,,\\
k'_{10} &=& k_5\,,\\
k'_{11} &=& k_{10}\,,\\
k'_{12} &=& k_9 -{1 \over 2}( 159 +288\eta +12\eta^2)\beta_3\,,\\
k'_{13} &=& k_8 -{1 \over 2}(144 +179\eta +40\eta^2)\beta_3 
-( 14 +6\eta)\rho_2 -(35 +15\eta)\rho_4\,,\\
k'_{14}&=& k_7 -{ 1\over 8}( 317 +105\eta)\eta\beta_3 
-{1 \over 2}( 84 +3\eta)\rho_4 -3\eta\rho_2 +2\chi_4 +7\chi_7\,,\\
k'_{15}&=& 8\chi_7\,,
\end{eqnarray}
\end{mathletters}
where $h_i$, $k_i$  are given by Eqs.~(\ref{sol4.5}) of the text 
and Eqs.~(\ref{4.5nrs})
remain the same.

\subsection{The 3.5PN and the 4.5PN gauge arbitrariness}

Finally it can be shown that all the arbitrary parameters in the reactive
solution may be absorbed in a choice of `gauge' of the form 
\begin{equation}
\delta {\bf x} = \frac{8}{5}\eta\frac{m}{r}
                 (f'_{2.5} + f'_{3.5} + f'_{4.5}) \dot r {\bf x}  +
(g'_{2.5}+g'_{3.5}+g'_{4.5}) r {\bf v} \,,
\label{deltax2}
\end{equation}
where  $ f'_{2.5}$ and $g'_{2.5} $ 
are given by Eqs.~(\ref{nsNGF}), (\ref{nsNGS}), while 
$f'_{3.5}$, $f'_{4.5}$, $g'_{3.5}$ and $g'_{4.5}$ have the form
\begin{eqnarray}
f'_{3.5} &=& \biggl [ P'_{21}v^4 +P'_{22}v^2{m \over r} +P'_{23}v^2\dot r^2
+P'_{24}{m \over r}\dot r^2 
+ P'_{25}\left({ m\over r}\right)^2 +P'_{26}\dot r^4\biggr]\,,\nonumber\\
g'_{3.5} &=& \biggl [ Q'_{21}v^4 +Q'_{22}v^2{m \over r} +Q'_{23}v^2\dot r^2
+Q'_{24}{m \over r}\dot r^2 
+ Q'_{25}\left({ m\over r}\right)^2 +Q'_{26}\dot r^4
\biggr ]\,,\nonumber\\
f'_{4.5}&=& \biggl [ P'_{41}v^6 +P'_{42}{m \over r}v^4 +P'_{43}v^4\dot r^2
+ P'_{44}v^2\left({ m\over r}\right)^2 
+P'_{45}v^2{m \over r}\dot r^2 \nonumber\\
&& +  P'_{46}v^2\dot r^4 + P'_{47}{m \over r}\dot r^4 
+P'_{48}\left({m \over r}\right)^2\dot r^2 
+P'_{49}\left({ m\over r}\right)^3 
+ P'_{410}\dot r^6 \biggr ]\,,\nonumber\\
g'_{4.5}&=& \biggl [ Q'_{41}v^6 +Q'_{42}{m \over r}v^4 +Q'_{43}v^4\dot r^2
+ Q'_{44}v^2\left({ m\over r}\right)^2 
+Q'_{45}v^2{m \over r}\dot r^2 \nonumber\\
&& +  Q'_{46}v^2\dot r^4 + Q'_{47}{m \over r}\dot r^4 
+Q'_{48}\left({m \over r}\right)^2\dot r^2 
+Q'_{49}\left({ m\over r}\right)^3 
+ Q'_{410}\dot r^6 \biggr ]\,.
\end{eqnarray}

At $3.5$PN we have
\begin{mathletters}
\begin{eqnarray}
P'_{21}&=& -{ 1\over 4}(1 -3\eta)\beta_3 -{ 1\over 4}(2 \rho_2 +\rho_4)\,,\\
P'_{22} &=& P_{21}+ { 1 \over 3}( 3 -10\eta)\beta_3 
+{ 1 \over 30}(20 \rho_2 +17\rho_4)\,,\\
P'_{23} &=& -{ 1 \over 4}\rho_4\,,\\
P'_{24} &=& P_{23} + { 1 \over 10}( 5\eta \beta_3 +\rho_4)\,,\\
P'_{25} &=& P_{22} +{ 1\over 12}( 2 +25\eta)\beta_3
 -{1 \over 3}(\rho_2 +\rho_4)\,,\\
P'_{26} &=& 0\,,\\
Q'_{21} &=& 0\,,\\
Q'_{22} &=& Q_{21} -{ 1 \over 2}(1 -3\eta)\beta_3
-{ 1\over 30}(10\rho_2 +7\rho_4)\,,\\
Q'_{23} &=& { 1\over 4}(1 -3\eta)\beta_3 +{1 \over 4}(2\rho_2 +\rho_4)\,,\\
Q'_{24} &=& Q_{23} -{1 \over 6}(3 -8\eta)\beta_3 
-{1 \over 30}(10\rho_2 +13\rho_4)\,,\\
Q'_{25}&=& Q_{22} -{ 1\over 12}(2 +25\eta)\beta_3 
+{1 \over 3}(\rho_2 +\rho_4)\,,\\
Q'_{26}&=& {1\over 4}\rho_4\,.
\end{eqnarray}
\end{mathletters}
Similarly at $4.5$PN we have
\begin{mathletters}
\begin{eqnarray}
P'_{41}&=& -{ 1\over 16}( 1 -9\eta +21\eta^2)\beta_3 
+{1 \over 4}( 1 -3\eta)\rho_2 +{ 1\over 8}(1 -3\eta)\rho_4\nonumber\\
&&
 -{1 \over 24}(12 \chi_2 +6\chi_4 +6\chi_7)\,,\\
P'_{42} &=& P_{41} -{ 1\over 840}(1155 -4817\eta +367\eta^2)\beta_3
-{1 \over 60}(130 -318\eta)\rho_2 
-{ 1\over 30}(53 -117\eta)\rho_4\nonumber\\
&&
+{ 1\over 105}(105\chi_2 +84\chi_4 +68\chi_7)\,,\\
P'_{43}&=& { 1\over 8}(1 -3\eta)\rho_4 -{1 \over 24}(6\chi_4 +4\chi_7)\,,\\
P'_{44} &=& P_{42} +{ 1\over 120}(420 -1917\eta +967\eta^2)\beta_3 
+ {1 \over 30}(55 -228\eta)\rho_2 
+{ 1\over 120}(220 -834\eta)\rho_4\nonumber\\
&&
-{ 1\over 15}( 15\chi_2 +14\chi_4 + 13\chi_7)\,,\\
P'_{45} &=& P_{43} -{ 1\over 140}\eta (47 +48\eta)\beta_3 
-{4 \over 5}\eta \rho_2 -{1 \over 20}(8 -17\eta)\rho_4 
+{1 \over 105}(21 \chi_4 +32\chi_7)\,,\\
P'_{46} &=& -{ 1\over 6}\chi_7\,,\\
P'_{47} &=& P_{46} -{ 27 \over 56}\eta (1 -3\eta)\beta_3
 -{ 1\over 4}\eta\rho_4 +{ 1\over 21}\chi_7\,,\\
P'_{48} &=& P_{45} +{ 1\over 600}(300 +4935\eta -1360\eta^2)\beta_3 
+{ 1\over 20}\eta ( 12\rho_2-\rho_4)-{ 1\over 15}(\chi_4 +2\chi_7)\,,\\
P'_{49} &=& P_{44} -{ 1\over 60}(92 +121\eta +309\eta^2)\beta_3 
+{ 1\over 15}(1 +52\eta)(\rho_2 +\rho_4) 
+{2 \over 5}(\chi_2 +\chi_4 +\chi_7)\,,\\
P'_{410} &=&0\,,\\
Q'_{41} &=& 0\,, \\
Q'_{42} &=& Q_{41}+ { 1\over 840}(105 -659\eta -347\eta^2)\beta_3
+{ 1\over 2}(1 -3\eta)\rho_2 +{ 7\over 20}(1 -3\eta)\rho_4\nonumber\\
&&
-{ 1\over 30}(10\chi_2 +7\chi_4) -{19\over 105}\chi_7\,,\\
Q'_{43} &=& { 1\over 16}( 1 -9\eta +21\eta^2)\beta_3
 -{1 \over 8}(1 -3\eta)(2\rho_2+\rho_4)
+{1 \over 24}( 12\chi_2 +6\chi_4 +4\chi_7)\,,\\
Q'_{44} &=& Q_{42} -{1 \over 240}(420 -1604\eta +1434\eta^2)\beta_3
-{1 \over 60}(80 -301\eta )\rho_2 
-{1 \over 30}(40 -131\eta)\rho_4\nonumber\\
&&
+{ 1\over 15}(10 \chi_2 +9\chi_4 +8\chi_7)\,,\\
Q'_{45} &=& Q_{43} + { 1\over 140}( 175 -639\eta +146\eta^2)\beta_3
+{ 1\over 30}(50 -99\eta)\rho_2 
+{1 \over 60}(94 -183 \eta)\rho_4\nonumber\\
&&
-{1 \over 21}( 14\chi_2 +14\chi_4 +12\chi_7)\,, \\
Q'_{46} &=& -{1 \over 8}( 1 -3\eta)\rho_4 
+{ 1\over 24}( 6\chi_4 +4\chi_7)\,,\\
Q'_{47} &=& Q_{46} + { 1\over 280}\eta (121 -363\eta)\beta_3
+{ 3\over 10}\eta \rho_2 +{1 \over 20}(5 -8\eta)\rho_4
-{ 1\over 10}\chi_4 -{26\over 105}\chi_7\,,\\
Q'_{48} &=& Q_{45} -{ 1\over 60}(135 -64\eta -11\eta)\beta_3
-{1 \over 60}(30 -119\eta)\rho_2 
-{1 \over 30}(15 -79\eta)\rho_4\nonumber\\
&&
+{ 1\over 15}(5 \chi_2 + 6\chi_4 +7\chi_7)\,,\\
Q'_{49} &=& Q_{44} + { 1\over 60}( 92 +121\eta +309\eta^2)\beta_3
-{ 1\over 15}(1 +52\eta)(\rho_2 +\rho_4) 
-{ 2\over 5}( \chi_2 +\chi_4 +\chi_7)\,,\\
Q'_{410} &=& { 1\over 6}\chi_7\,.
\end{eqnarray}
\end{mathletters}
In the above, the  $P_{ab}$ and $Q_{ab}$  are given by
Eqs.~(\ref{3.5PNgauge}) and (\ref{2PNgauge})
of the text.

To conclude: the far-zone flux formulas and the balance equations by
themselves do not constrain the reactive acceleration to be a power
series in $m_1$ and $m_2$, or equivalently nonlinear in the total mass
$m$, as assumed in the paper, following IW. They are also consistent
with the more general form of the reactive acceleration discussed in
this appendix.

\newpage
\begin{table}
\caption{ Comparison of four Alternative Schemes : IW21, IW22 (Minimal) 
IW11 and IW00. 
N denotes the order of approximation, NV the number of
variables, NC the number of constraints coming from balance equations, 
ND the number of degenerate equations, NI the number of independent
equations and NA the number of arbitrary parameters determining the
solution. In the NV column, $a+b+c$ means $a$ variables of reactive
acceleration, $b$ in energy ambiguity and $c$ in angular momentum
ambiguity.}
\label{tab:altscheme}
\begin{tabular}{cccccc}
N & NV  & NC & ND & NI & NA  \\
\tableline 
\multicolumn{6}{c}{IW21: IW Scheme} \\ 
\tableline        
2.5PN &6+3+3& 12& 2& 10& 2 \\
3.5PN &12+6+6 & 20& 2 &18 & 6 \\
4.5PN &20+10+10&  30 & 2  &28 & 12 \\ 
\tableline
\multicolumn{6}{c}{IW22: Minimal Scheme} \\ 
\tableline
2.5PN &6+3 +1& 9& 1& 8& 2 \\
3.5PN &12+6+3 & 16& 1 &15 & 6 \\
4.5PN &20+10+6& 25  & 1  &24 & 12 \\ \tableline
\multicolumn{6}{c}{IW11 Scheme} \\ 
\tableline
2.5PN &6+6+3&16 & 3& 13& 2 \\
3.5PN &12+10+6 &25& 3 &22 & 6 \\
4.5PN &20+15+10& 36  & 3  &33 & 12 \\\tableline
\multicolumn{6}{c}{IW00 Scheme} \\ 
\tableline
2.5PN &6+10+6& 25&5&20&2\\\tableline
\end{tabular}
\end{table} 
\begin{table}
\caption{Comparison of the four Alternative Schemes:  
EJ21, EJ22, EJ11, and EJ00 at 2.5PN level. The notation is as in Table 
\protect\ref{tab:altscheme}. In the NC column, $a+b$ indicates that $a$
constraints  arise from energy balance and $b$ from angular momentum
balance.}
\label{tab:altascheme}
\begin{tabular}{cccccc}
Scheme & NV  & NC & ND & NI & NA  \\
\tableline 
EJ22 &12+3+1& 10+6& 2& 14& 2 \\
EJ21 &12+3 +3&10+6 & 1& 15& 3 \\
EJ11 &12+6+3&10+6 & 1& 15& 6 \\
EJ00 &12+10+6&15+10 &3&22&6\\
\tableline
\end{tabular}
\end{table} 


\begin{references}
\bibitem{newnotn} For consistency of notation we have renamed the variables
in IW in this paper. The following is a  `map' between the IW
variable and the corresponding notation used here, denoted as 
`old' $\to $ `new'. \\
$A_{5\over 2} \to A_{2.5}, B_{5\over 2} \to B_{2.5},
A_{7\over 2} \to A_{3.5}$,
$B_{7\over 2} \to B_{3.5}$, $ c_i \to f_i$, $d_i \to g_i$,
$\delta_i\to\xi_i$, $\varepsilon_i\to\rho_i$, where $i=1\ldots6$,
$\gamma \to \alpha_1, \delta \to \alpha_2, \beta \to \alpha_3$
and $ \kappa \to \beta_1, \alpha \to \beta_2, \epsilon \to \beta_3 $.
\bibitem{nonlin} The combination $\eta(m/r^2)(m/r) $ chosen as prefactor 
in the IW 
ansatz for acceleration is proportional to $m_1m_2$ or $G^2$
while the combination $\eta (m/r^2)$ chosen in the Appendix is
proportional to $m_1m_2 (m_1+m_2)^{-1}$. Unlike the former choice,
the latter is not a power series in $m_1$ and $m_2$ and hence
while the former is the `physically relevant solution'
the latter is of `mathematical interest'.
\bibitem{Orbit}
It is also possible to obtain $\dot r$ and ${\dot \omega} $ using 
2PN representation of orbits \cite{DS88,SW93,W95,WR95}. 
Here one writes down `r' in harmonic coordinates in terms of
conserved energy E. To compute $\dot r $ one then requires to 
calculate $ \dot E $ using,  acceleration to $O(\epsilon^{4.5})$.
\end{references}
\end{document}